# Connection between continuous and digital n-manifolds and the Poincaré conjecture.

Alexander V. Evako
NPK Novotek.
Volokolamskoe Sh. 1, kv. 157, 125080, Moscow, Russia
Tel/Fax: 499 158 2939, e-mail: evakoa@mail.ru.

Abstract.
We introduce LCL covers of closed n-dimensional manifolds by n-dimensional disks and study their properties. We show that any LCL cover of an n-dimensional sphere can be converted to the minimal LCL cover, which consists of 2n+2 disks. We prove that an LCL collection of n-disks is a cover of a continuous n-sphere if and only if the intersection graph of this collection is a digital n-sphere. Using a link between LCL covers of closed continuous n-manifolds and digital n-manifolds, we find conditions where a continuous closed three-dimensional manifold is the three-dimensional sphere. We discuss a connection between the classification problems for closed continuous three-dimensional manifolds and digital three-manifolds.

Key words: Digital model, digital surface, space, manifold, graph, dimension, digital topology, cover,

1. Introduction

A digital approach to geometry and topology plays an important role in analyzing n-dimensional digitized images arising in computer graphics as well as in many areas of science including neuroscience, medical imaging, industrial inspection, geoscience and fluid dynamics. Concepts and results of the digital approach are used to specify and justify some important low-level image processing algorithms, including algorithms for thinning, boundary extraction, object counting, and contour filling [1-3,5-9,13,17,22,23].
We use an approach in which a digital n-surface (digital normal n-dimensional space) is considered as a simple undirected graph of a specific structure. Properties of n-surfaces were studied in [5-9]. Paper [8] analyzes a local structure of the digital space $Z^n$. It is shown that $Z^n$ is an n-surface for all n>0. In paper [9], it is proven that if A and B are n-surfaces and A⊆B, then A=B. This paper presents conditions which guarantee that every digitization process preserve certain topological and geometrical properties of continuous closed two-surfaces. In papers [5-7], X. Daragon, M. Couprie and G. Bertrand introduce and study the notion of frontier order, which allows defining the frontier of any object in an n-dimensional space. In particular, they investigate a link between abstract simplicial complexes, partial orders and n-surfaces. In the framework of abstract simplicial complexes, they show that n-dimensional combinatorial manifolds are n-surfaces and n-surfaces are n-dimensional pseudomanifolds and that the frontier order of an object is the union of disjoint (n-1)-surfaces if the order to which the object belongs is an n-surface.
A digital n-manifold which we regard in this paper is a special case of a digital n-surface.
It seems desirable to consider properties of digital n-manifolds in a fashion that more closely parallels the classical approach of algebraic topology in order to find out, how far the fundamental distinction between continuous and digital spaces due to different cardinality restricts a direct modification of continuous tools to digital models on one hand and how effectively the digital approach can be applied to solve classical topology problems on the other hand. As an example, we consider the Poincaré conjecture about the characterization of the 3-dimensional sphere amongst 3-dimensional manifolds.

The review of some of the major results obtained in an attempt to prove the Poincaré conjecture may be found in [15]. Recently, three groups have presented papers that claim to complete the proof of the Poincaré conjecture. The results of these papers are based upon earlier papers by G. Perelman [19-21].

In May 2006, B. Kleiner and J. Lott posted a paper [16] on the Arxiv. They claim to fill in the details of Perelman's proof of the Geometrization conjecture.

In June 2006, H-D. Cao and X-P. Zhu published a paper [4] claiming that they give a complete proof of the Poincaré and the geometrization conjectures.

In July 2006, J. Morgan and G. Tian posted a paper [18] on the Arxiv in which they claim to provide a detailed proof of the Poincaré Conjecture.

Our approach to the characterization of the 3-dimensional sphere amongst 3-dimensional manifolds is different from previous attempts. It is based on the connection between LCL covers of closed n-manifolds and digital n-manifolds.

In section 2, we describe computer experiments which provide a reasonable background for introducing digital spaces as simple graphs. Then we remind basic definitions and results related to digital n-dimensional spaces (n-spaces) (section 3). In sections 4, we study properties of digital n-disks and n-spheres, which are similar to properties of their continuous counterparts. We introduce disk transformations of digital n-manifolds, which retain their basic features. It is proven that a digital n-sphere converts into the minimal one by disk transformations and that a digital n-sphere without a point is homotopic to a point. In sections 5, we study properties of compressed digital n-manifolds. In section 6, we introduce LCL collections of n-dimensional continuous disks. We consider a decomposition of a closed continuous n-manifold to an LCL union of n-disks and study properties of the cover. We find conditions where an LCL collection is a cover of a continuous n-dimensional sphere. We prove that a given continuous closed n-manifold is an n-dimensional sphere if any LCL cover of this manifold can be converted to the minimal one consisting of 2n+2 elements by the merging of n-disks. The results of sections 4, 5 and 6 are based on results obtained in [10] and [11]. We find a link between intersection graphs of LCL covers of continuous closed n-manifolds and digital n-manifolds (section 7). In sections 8 and 9 apply obtained results to find conditions including Poincaré conjecture about the characterization of continuous 2- and 3-dimensional spheres amongst closed continuous 2-, 3-dimensional manifolds. Finally, we discuss ways, which can help in treating the classification problem for closed 3-dimensional manifolds.

Throughout the paper, by a continuous n-manifold, we mean a closed (compact and without boundary) path-connected n-manifold and digital spaces all have a finite amount of points.

2. Computer experiments as the basis for digital spaces.

An important feature of this approach to the structure of digital spaces is that it is based on computer experiments whose results can be applied to computer graphics and animations.

The following surprising fact is observed in computer experiments modeling deformations of continuous surfaces and objects in 3-dimensional Euclidean space $R^3$ [14]. Suppose that $S_1$ is a surface (in general, a 1- 2- or 3-dimensional object) in $R^3$. Tessellate $R^3$ into a set of unit cubes, pick out the family $M_1$ of unit cubes intersecting $S_1$ and construct a digital space $D_1$ corresponding to $M_1$ as the intersection graph of $M_1$ [12]. Then reduce the size of the cube edge from 1 to 1/2 and using the same procedure, construct $D_2$. Repeating this operation several times, we obtain a sequence of digital spaces $D=\{D_1,D_2,...D_n,...\}$ for $S_1$. It is revealed that there exists a number p such that for all m and n, m>p, n>p, $D_m$ and $D_n$ can be turned from one to the other by four kinds of transformations called contractible. For example, a digital model of a closed simple curve can be converted to the digital minimal 1-sphere (fig. 4.3), a digital model of a segment, a piece of plane or a 3-dimensional cube converts to a point and so on. It is reasonable to assume that digital spaces contain topological, and perhaps geometric characteristics of continuous surfaces.

3. Preliminaries.

By a digital space G, we mean a simple graph G=(V,W) with a finite or countable set of points $V=\{v_1,v_2,...v_n,...\}$, together with a set of edges $W = \{(v_pv_q),....\} \subseteq V \times V$, provided that $(v_pv_q)=(v_qv_p)$ and $(v_pv_p) \notin W$ [12]. Points $v_p$ and $v_q$ are called adjacent if $(v_pv_q) \in W$. We use the notations $v_p \in G$ and $(v_pv_q) \in G$, if $v_p \in V$ and $(v_pv_q) \in W$ respectively if no confusion can result. A graph K is called complete if every two points of K are adjacent. A graph $v \oplus G$ is a graph where point v is adjacent to all points of graph G. A graph $v \oplus G$ is called the cone of a graph G. $H=(V_1,W_1)$ is a subspace of G=(V,W) if $V_1 \subseteq V$ and points $v_p,v_q \in H$ are adjacent in H if and only if they are adjacent in G. In other words, H is the induced subgraph of a graph G. Let G be a digital space and v be a point of G. The subspace U(v) containing v as well as all its neighbors is called the ball of point v in G. The subspace O(v)=U(v)-v containing only neighbors of v is called the rim of point v in G.
The subspace $O(v_1,v_2,v_3,....v_P,)$ formed by the intersection of $O(v_1)$, $O(v_2)$, $O(v_3)$, ... $O(v_P)$ is called the joint rim of points $v_1$, $v_2$, $v_3$,.... $v_P$. Digital spaces can be transformed from one into another in variety of ways. Contractible transformations of digital spaces [14] seem to play the same role in this approach as a homotopy in algebraic topology.
Definition 3.1.
The family $T=(K(1),G_1,G_2,G_3,...G_n,...)$ of graphs $G_1, G_2, G_3,...G_n,...$ is called contractible if:
One-point graph K(1) belongs to T, $K(1) \in T$.
Any graph G of T can be obtained from any other graph of T by following transformations:
Deleting of a point v. A point v of a graph G can be deleted, if $O(v) \in T$.
Gluing of a point v. If a subgraph H of a graph G belongs to family T, $H \in T$, then a point v can be glued to the graph G in such a manner that O(v)=H.
Deleting of an edge (v,u). An edge (v,u) of a graph G can be deleted if the joint rim O(v,u) belongs to T, $O(v,u) \in T$.
Gluing of an edge (v,u). Let two points v and u of a graph G be non-adjacent. An edge (v,u) can be glued to G if the joint rim O(v,u) belongs to T, $O(v,u) \in T$.
Any graph belonging to T is called contractible. Some contractible graphs are depicted in fig. 3.1.

Definition 3.2.

The following transformations of a graph G are said to be contractible [14].

Deleting of a point v. A point v of a graph G can be deleted, if $O(v) \in T$.

Gluing of a point v. If a subgraph H of a graph G belongs to family T, $H \in T$, then a point v can be glued to the graph G in such a manner that $O(v)=H$.

Deleting of an edge (v,u). An edge (v,u) of a graph G can be deleted if the joint rim $O(v,u)$ belongs to T, $O(v,u) \in T$.

Gluing of an edge (v,u). Let two points v and u of a graph G be non-adjacent. An edge (v,u) can be glued to G if the joint rim $O(v,u)$ belongs to T, $O(v,u) \in T$.

Definition 3.3.

If a graph G can be obtained from a graph H by a sequence $f_1, f_2, \ldots f_n$ of contractible transformations, then we say that G is homotopic to H and write $G = f_n \ldots f_2 f_1 H$, $G \sim H$.

Homotopy is an equivalence relation among digital spaces. Contractible transformations retain the Euler characteristic and homology groups a graph.

Subgraphs A and B of a graph G are called separated or non-adjacent if any point in A is not adjacent to any point in B.

The join $G \oplus H$ of two separated spaces $G=(X,U)$ and $H=(Y,W)$ is the space that contains G, H and edges joining every point in G with every point in H [8]. In graph theory, this operation is also called the join of two graphs [12]. Remind the isomorphism of digital spaces. Note that the isomorphism of digital spaces is the isomorphism of graphs [12]. A digital space G with a set of points $V=\{v_1,v_2,...v_n\}$ and a set of edges $W=\{(v_p,v_q),....\}$ is said to be isomorphic to a digital space H with a set of points $X=\{x_1,x_2,...x_n\}$ and a set of edges $Y=\{(x_p,x_q),....\}$ if there exists one-one onto correspondence f: $V \to X$ such that $(v_i,v_k)$ is an edge in G iff $(f(v_i),f(v_k))$ is an edge in H. Map f is called an isomorphism of G to H. We write G=H to denote the fact that there is an isomorphism of G to H. Let G and H be digital spaces and A and B be their subspaces, $A \subseteq G$, $B \subseteq H$. If A and B are isomorphic, then the space G#H obtained by identifying points in A with corresponding points in B is said to be the connected sum of G and H over A (or B).

Let $W=\{X_1,X_2,....X_n,...\}$ be a finite or countable family of sets. Then graph G(W) with points $\{v_1,v_2,....v_n,...\}$, where $v_k$ and $v_i$ are adjacent whenever $X_k \cap X_i \neq \emptyset$ is called the intersection graph of family W [12].

Definition 3.4.

A normal digital 0-dimensional space is a disconnected graph $S^0(a,b)$ with just two points a and b. For n>0, a normal digital n-dimensional space is a nonempty connected graph $G^n$ such that for each point v of $G^n$, O(v) is a normal finite digital (n-1)-dimensional space. [8].

The normal digital 0-dimensional space is called the normal digital 0-dimensional sphere.

Proposition 3.1 [8].

Let $G^n$ and $H^m$ be normal n- and m-dimensional spaces. Then their join $G^n \oplus H^m$ is a normal (n+m+1)-dimensional space.

Proposition 3.2.

Let $G^n$ be a normal n-dimensional space and $K\{v_1,v_2,...v_k\}$ $1 \leq k \leq n$ is a complete subspace of $G^n$. (every two points are adjacent) determined by points $\{v_1,v_2,...v_k\}$. Then the joint rim $O(v_1,v_2,...v_k)$ of these points (the mutual adjacency set) is a normal (n-k)-dimensional space [8].

Proposition 3.3.

If $G^n$ and $H^n$ are normal n-dimensional spaces and $H^n$ is a subspace of $G^n$, then $H^n=G^n$ [9].

Proposition 3.4 [14].
- The cone $v\oplus G$ of any space G is a contractible space.
- Let G be a contractible graph and H be its contractible subgraph. Then G can be converted into H by contractible deleting of points in any suitable order.
- Let G be a contractible graph. Then for any point v belonging to G, subgraphs O(v) and G-v are homotopic and G-v can be converted into O(v) by contractible deleting of points in any suitable order.

Proposition 3.5.
Let $G\{v_1,v_2,...v_t\}$ and $F=\{G_1,G_2,....G_n\}$ be a graph and a family of its non-empty subgraphs $G_k\{v_{k(i)}\}$, k=1,2,...n, of G with the following properties.
- Family F is a cover of G, $G=G_1\cup G_2\cup....G_n$.
- Any $G_k$, k=1,2,...n is contractible.
- From condition $G_{p(i)}\cap G_{p(k)}\neq\varnothing$, i,k=1,2,...m, it follows that $G_{p(1)}\cap G_{p(2)}\cap...\cap G_{p(m)}\neq\varnothing$.
- If $G_{p(1)}\cap G_{p(2)}\cap...\cap G_{p(m)}\neq\varnothing$, then $G_{p(1)}\cap G_{p(2)}\cap...\cap G_{p(m)}$ is contractible.
- For any point v of G, there exists subgraph $G_k$ such that the ball U(v) of point v belongs to $G_k$, $U(v)\subseteq G_k$.

Then the intersection graph G(F) of family F is homotopic to graph G. [9].

4. Properties of digital n-spheres and n-disks.

Since in this and next sections section, we consider only digital spaces, we will use the word space for digital space if no confusion can result. In order to make this section self-contained we will use the necessary information from paper [10].

A 0-ball and a 0-disk is a single point v, a 0-sphere $S^0(a,b)$ is a disconnected graph with just two points a and b. A 1-sphere $S^1$ is a connected graph such that for each point v of $S^1$, the rim O(v) of v is a 0-sphere $S^0$ (fig. 4.1). A 1-disk $D^1$ is a connected graph $S^1$-v obtained from a 1-sphere $S^1$ by the deleting of a point v.

Lemma 4.1.
The minimal 1-sphere $S^1_{min}$ consists of four points, $S^1_{min}=S^0(a,b)\oplus S^0(c,d)$.
The minimal 1-disk $D^1_{min}$ consists of three points, $D^1_{min}=a\oplus S^0(b,c)$
Any 0-sphere $S^0$ belonging to a 1-sphere $S^1$ divides $S^1$ into two separated parts. If $S^1$ is a 1-sphere and $S^0$ is a 0-sphere belonging to $S^1$, then $S^1$ is the union $A\cup S^0\cup B$ where subspaces A and B are separated and subspaces $A\cup S^0$ and $B\cup S^0$ are 1-disks.
The properties are checked directly.
To define n-disks and n-spheres, we will use a recursive definition. Suppose that we have defined k-disks and k-spheres for dimensions $1\leq k\leq n-1$.

Definition 4.1.
( a ) A connected space N is called an n-manifold if the rim of any point of N is an (n-1)-sphere.
( b ) A connected space N is called an n-manifold with boundary $\partial N$, if there exists an n-manifold M and a point v in M such that M-v is isomorphic to N. The subspace O(v) is called the boundary of N, the subspace $IntN=N-\partial N$ is called the interior of N. Obviously, the boundary $\partial N$ of N is an (n-1)-sphere.

Remark 4.1.
Any n-manifold is a normal n-dimensional space, but a normal n-dimensional space in not necessarily an n-manifold. For example, the join $A=S^0(a,b)\oplus P^2$ of a 0-sphere $S^0$ and a 2-dimensional projective plane $P^2$ is a normal 3-dimensional space, but A is not a 3-manifold because the rims of points a and b are not 2-spheres.

Definition 4.2.
( a ) A connected space D is called an n-disk if it has the following properties:
- D is a contractible graph (that is D can be converted to a point by contractible transformations).
- D can be represented as the union $D=\partial D\cup IntD$ of two non-empty subspaces such that if a point v belongs to $\partial D$, then the rim O(v) of v is an (n-1)-disk and if a point v belongs to IntD, then O(v) is an (n-1)-sphere.
- The boundary $\partial D$ of D is an (n-1)-sphere.

( b ) Let D and C be n-disks such that $\partial D$ and $\partial C$ are isomorphic, $\partial D=\partial C$. The space D#C obtained by identifying each point in $\partial D$ with its counterpart in $\partial C$ is called an n-sphere (fig. 4.2).

Obviously, S is the connected sum of D and C over $\partial D$.

Remark 4.2.
Suppose that N is an n-manifold and v is a point belonging to N. Then according to definition 4.2 and proposition 3.4, the ball U(v) of point v is an n-disk with the boundary O(v), $\partial U=O(v)$, and the interior containing only one point v, IntU(v)=v. Let us call U(v) an n-ball.

Definition 4.3.
The join $S^n_{min}=S^0_1\oplus S^0_2\oplus\ldots S^0_{n+1}$ of (n+1) separated copies of the 0-sphere $S^0$ is called the minimal n-sphere [9]. The join of the minimal (n-1)-sphere $S^{n-1}_{min}$ and a point v not belonging to $S^{n-1}_{min}$ is called the minimal n-disk, $D_{min}=v\oplus S^{n-1}_{min}$ (fig. 4.3).

Remark 4.3.
As it follows from definitions 4.2 and 4.3, the minimal n-sphere is an n-sphere and an n-manifold.

Lemma 4.2.
( a ) Any n-sphere is an n-manifold.
( b ) Let S be an n-sphere and v be a point of S. Then S-v obtained from S by the deleting of point v is an n-disk (fig. 4,2, 4.4).
Proof.
To prove ( a ), use induction. For n=1, the lemma is plainly true. Assume that the lemma is valid for dimensions n<k. Suppose that n=k. By definition 4.2, S is the connected sum D#C of n-disks D and C over $\partial D$. Let a point v belongs to $\partial D$. Then the rim O(v) in S is the union $O(v)=A\cup B$, where $A=O(v)\cap D$ is the rim of v in D and $B=O(v)\cap C$ is the rim of v in C. By definition 4.2, A and B are (n-1)-disks with isomorphic boundaries $\partial A$ and $\partial B$. By construction, O(v)=A#B is the connected sum of A and B over $\partial A$. Therefore, O(v) is an (n-1)-sphere by the induction. The rims of all points belonging to IntD and IntC are (n-1)-spheres. According to definition 4.1, S is an n-manifold.
To prove ( b ), note that S-v is an n-manifold with boundary by ( a ). Therefore, we have to prove that S-v is contractible. Let us use a double induction.
For n=1, the lemma is plainly true. Assume that the lemma is valid for dimensions n<k. Suppose that n=k.
Note that for $S=S^n_{min}$, the lemma is obvious. Assume that the lemma is valid for S with a number of points |S|=r≤t. Let r=t+1. By definition 4.2, S is the connected sum D#C of n-disks D and C over ∂D. With no loss of generality, suppose that a point v belongs to the interior of D, v∈IntD, and |IntC|>1. Suppose that a point x is separated from S, connect point x with any point belonging to C and delete all points belonging to IntC. Obviously, this is a sequence $\{g_1…g_m\}$ of contractible transformations and the obtained space N=S+v-IntC is homotopic to S. By construction, N is a connected sum, N=E#D, where E=v⊕∂C, |N|<t+1. Therefore, N is the n-sphere by definition 4.2. Hence, N-v=F is an n-disk by the assumption i.e., F is a contractible space. Obviously, S-v can be converted to F=N-v by the same sequence $\{g_1…g_m\}$ of contractible transformations. Therefore, S-v is homotopic to F=N-v. Since F=N-v is contractible, then S-v is contractible. □

Figure 4.4 shows a 2-sphere S and a 2-dimensional projective plane P. S-v is a 2-disk, which is homotopic to a point. P-v is not a 2-disk, it is homotopic to a 1-sphere S.

Lemma 4.3.
Let S be an n-sphere and D be an n-disk belonging to S. Then S-IntD is an n-disk.
Proof.
We have to prove that S-IntD is a contractible space and an n-manifold with boundary. Note that if D is the ball U(v) of a point v, then IntD=v and C=S-IntD=S-v is an n-disk by lemma 4.2.
( a ) Let us prove that S-IntD is a contractible space. Suppose that IntD contains more than one point. Take disk D separately from S and take an n-disk E=v⊕∂D separated from D and S. Then a space D#E over ∂D is an n-sphere by definition 4.2. Let a point u belong to IntD. Then F=D#E-u is an n-disk by lemma 4.2. Therefore, F is a contractible space. By proposition 3.4, F-v converts into O(v)=∂D by the contractible deleting of points. Therefore, S-v converts into S-IntD by the same contractible deleting of points. Hence C=S-IntD is a contractible space.
( b ) The proof that S-IntD is an n-manifold with boundary is similar to the poof of assertion ( a ) in lemma 4.2 an is omitted. □

The following corollary is an easy consequence of lemma 4.2.

Corollary 4.1.
Let N be an n-manifold with boundary ∂N. Then
- ∂N is an (n-1)-sphere.
- If a point v belongs to ∂N, then O(v) is an (n-1)-disk.
- If a point v belongs to IntN, then O(v) is an (n-1)-sphere.

Lemma 4.4.
Let M be an n-manifold. If the rim O(v) of any point v belonging to M is the minimal (n-1)-sphere, then M is the minimal n-sphere.
Proof.

Suppose that the rim O(v) of a point v of M is the minimal (n-1)-sphere with points $\{v_1,v_2,\ldots v_{2n}\}$ where point $v_{2k+1}$ is adjacent to all points except $v_{2k+2}$, k=0,1,…n-1. Consider the rim of point $v_1$. $O(v_1)$ contains points $\{v_3,v_4,\ldots v_{2n},v,u\}$ where a point u is adjacent to all points except v. Consider the rim of point $v_3$. $O(v_3)$ contains points $\{v_1,v_2,v_5,v_6,\ldots v_{2n},v,u\}$. Since points u and v are non-adjacent, then u must be adjacent to $v_2$. Therefore, subspace G of M consisting of points $\{v_1,v_2,\ldots v_{2n},v,u\}$ is the minimal n-dimensional sphere. Since G⊆M, then G=M according to proposition 3.3.
☐

Definition 4.4.
( a ) Let N be an n-manifold, D be an n-disk belonging to N and v be a point belonging to IntD. We say that IntD is replaced with point v (or D with v⊕∂D) if the space M=N+v-IntD is obtained by deleting from N points belonging to IntD except point v and joining point v with any point in ∂D. Let denote this transformation by (D,v) (fig. 4.5).
( b ) Let N be an n-manifold and v be a point belonging to N. Let D be an n-disk not belonging to N and such that O(v) is isomorphic to ∂D. We say that point v is replaced with IntD (or the n-ball U(v) is replaced with D) if the space M=N-v+IntD is obtained by identifying any point in ∂D with its counterpart in O(v) and deleting point v from N. Let denote this transformation by (v,D).

In definition 4.4(a), we can take a point v not belonging to N, delete all points belonging to IntD and connect v with all points belonging to ∂D. The result will be the same.
We call replacings (a) and (b) disk transformations or d-transformations. (D,v)- and (v,D)-transformations can be considered as the merging of points belonging to IntD into a point v and the splitting of a point v into a collection of points belonging to IntD.
We say that points $v_1,v_2,\ldots v_k$ can be merged (into one point) if there is an n-disk D belonging to N such that $v_1,v_2,\ldots v_k$ belong to IntD.
In fact, both operations are the replacings of n-disks by n-disks. The replacing of n-disks in an n-manifold N is an application of contractible transformations [14] of digital spaces to n-manifolds. d-Transformations are represented by a sequence of contractible transformations of digital spaces that retain such properties of digital spaces as the Euler characteristic and the homology groups.

Lemma 4.5.
Any d-transformation is a sequence of contractible transformations.
Proof,
To prove that the d-transformation (D,v) is a sequence of contractible transformations, suppose that N is an n-manifold, D is an n-disk belonging to N and a point v does not belong to N. Since D is a contractible space, then point v can be glued to N in such a manner that O(v)=D according to definition 3.2. This is a contractible transformation converting N into N+v. Suppose that a point u belongs to IntD. Denote by O(u) the rim of u in N. Then the rim A(u) of u in N+v is the join v⊕O(u) of v and O(u), A(u)=v⊕O(u). According to proposition 3.4, A(u) is a contractible space. Therefore, point u can be deleted from N+v. In the same way, any point belonging to IntD can be deleted from N+v. Hence, the obtained space M=N+v-IntD is homotopic to N.
In the same way, it is easy to prove that the d-transformation (v,D) is too a sequence of contractible transformations. □

Definition 4.5.
 n-Manifolds M and N are called equivalent if one of them can be obtained from the other by a sequence of d-transformations.

Lemma 4.6.
An n-sphere S is equivalent to the minimal n-sphere $S_{min}$.
Proof.
Suppose that S=D#C is the connected sum of n-disks D and C and points v and u are separated and not belonging to S. Replace IntD and IntC by points v and u according to definition 4.4. Then S converts into the n-sphere $S^0(v,u)\oplus\partial D$. Since $\partial D$ is an (n-1)-sphere, then for the same reason as above, $\partial D$ can be converted to $S^0(a,b)\oplus\partial E$ where $\partial E$ is an (n-2)-sphere. Hence, S can be turned in $S^0(v,u)\oplus S^0(a,b)\oplus\partial E$. Repeat the above transformations until we obtain the minimal n-sphere. □

Lemma 4.7.
d-Transformations convert an n-manifold into an n-manifold.
Proof.
Let N be an n-manifold N and let M=gN be a space obtained from N by a d-transformation g. We have to prove that for any point v of M, the rim O(v) is an (n-1)-sphere. Let D be an n-disk belonging to N. Suppose that M=N+v-IntD is obtained by connecting a point v belonging to IntD with any point in $\partial D$ and deleting from N all points belonging to IntD except v. We have to prove that the rim of any point of M is an (n-1)-sphere. Let a point x belong to $\partial D$, B(x) be the rim of x in M, O(x) be the rim of x in N and A(x)=O(x)∩D be the rim of point x in D. Since x is a boundary point of D, then A(x) is an (n-1)-disk according to corollary 4.1. Since O(x) is an (n-1)-sphere and A(x)⊆O(x), then C(x)=O(x)-IntA(x) is an (n-1)-disk by lemma 4.3. Obviously, $\partial C(x)=\partial A(x)$. Then B(x) is the connected sum of the n-disk C(x) and the n-disk $E=v\oplus\partial C(x)$ over $\partial C(x)$. Hence, B(x)=C(x)#E is the (n-1)-sphere by definition 4.2. If a point y does not belong to D, then its rim does not change and remains an (n-1)-sphere. The rim of point v is an (n-1)-sphere $\partial D$. Therefore, M is an n-manifold by definition 4.1.
Suppose that the boundary $\partial D$ of an n-disk D is isomorphic to the rim O(v) of some point v of N. In the same way as above, we can prove that M=N-v+IntD obtained by identifying any point in $\partial D$ with its counterpart in O(v) and deleting point v is an n-manifold. □

Lemma 4.8.
d-Transformations convert an n-sphere into an n-sphere.
Proof.
Let N=gS be the space obtained from an n-sphere S by an d-transformation g. We have to prove that N can be represented as the connected sum of n-disks A and B over $\partial A$.
Suppose that S is an n-sphere and v is a point belonging to S. Let D be an n-disk separated from S such that $\partial D$ is isomorphic to O(v). Suppose that N=gS=S-v+IntD is the space obtained by identifying any point in $\partial D$ with its counterpart in O(v) and deleting point v. Clearly, N is the connected sum of A=S-v and D over $\partial D$. Since A is an n-disk by lemma 4.2, then N=A#D is an n-sphere.

Suppose that S is an n-sphere and D be an n-disk belonging to S and a point v be separated from S. Let the space N=S+v-IntD is obtained by joining point v with any point in ∂D and deleting from S points belonging to IntD according to definition 4.4. Then N is the connected sum Of A=S-IntD and B=v⊕∂D. Since A is an n-disk by lemma 4.3 and B is an n-disk by definition 4.2, then N=A#B is an n-sphere by definition 4.2. □

The following theorem summarizes the previous results.

Theorem 4.1.
- d-Transformations convert an n-manifold into an n-manifold.
- d-Transformations convert an n-sphere into an n-sphere.
- An n-manifold M is an n-sphere if and only if M is equivalent to the minimal n-sphere $S_{min}$.
- An n-manifold M is an n-sphere if and only if there is a point v such that M-v is an a disk.

d-Transformations of an n-manifold M can change m-spheres and m-disks belonging to M, m<n. However, if a d-transformation f of M generates a d-transformation $f_N$ of an m-manifold N belonging to M, then we can say that N converts into $f_N$N belonging to fM.

Remark 4.4.
Suppose that M is an n-manifold M, S is an m-sphere belonging to M, C is an m-disk belonging to M, m<n. We say that S is embedded in D if there is an n-disk D such that S⊆D. If S∩IntD≠∅, then after the merging of all interior points of D into a point v, S is collapsed into a set of points which is not an m-sphere. If S⊆∂D, then after the merging of all interior points of D into an point v, S belongs to the boundary of an n-disk $D_1$ with only one interior point v (fig, 4.6).
We say that C is embedded in D if there is an n-disk D such that IntC⊆IntD (and C⊆D). If ∂C∩IntD≠∅, then after the merging of all interior points of D into a point v, D is collapsed into a set of points, which is not an m-disk. If ∂C⊆∂D, then after the merging of all interior points of D into a point v, C converts into an m-disk v⊕∂C with only one interior point v.

An m-sphere S is not necessarily the boundary of some (m+1)-disk D. For example, a 1-sphere S in a 3-manifold M may be a knot for which there exist no 2-disk D in M such that S=∂D.

Problem 4.1.
There is an open problem: Suppose that an n-manifold M is homotopic to an n-manifold N (M can be turned into
N by contractible transformations). Does it follow that M and N are equivalent?
This problem is linked with a similar problem arising in the study of LCL covers of closed continuous n-manifolds.

5. Compressed spaces.

Although in this section we deal with n-manifolds, most of the results can be applied to n-spaces in which the rim of a point is not necessarily an (n-1)-sphere.
As we have already mentioned, the main difference between digital and continuous n-manifolds is that a digital n-manifold has a finite or countable number of points while a continuous n-manifold has the cardinality of the continuum. If a digital n-manifold has a finite amount of points, it can be reduced by d-transformations while it is impossible for continuous spaces. This is essential for our further study because n-manifolds with a small number of points are easier to analyze. In the rest of the paper, we consider n-manifolds with n>0.

Definition 5.1.
An n-manifold N is called compressed if any n-disk D in N is the ball of some point v, D=U(v)
Fig. 4.3 and 4.4 show compressed minimal 1-, 2- and 3-spheres, a non-compressed 2-sphere S and a compressed 2-dimensional projective plane P. In S, the union of balls U(v)∪U(u) is a 2-disk.

Lemma 5.1.
Any n-manifold N can be compressed by d-transformations.
Proof.
Note first, that the ball U(v) of any point v of N is an n-disk. Take an n-disk D belonging to N and different from the ball of any point of N. Therefore, D contains more than one interior point. Introduce connections between a point v belonging to IntD and all points belonging to ∂D and delete all points belonging to IntD except for v. Then N moves to an n-manifold M equivalent to N. Repeat this procedure until any n-disk is the ball of some point. If N contains the finite number of points, the number of replacings is finite. This completes the proof. □
In the following lemma, we prove some properties of compressed n-manifolds which will be used further.

Lemma 5.2.
If N is a compressed n-manifold, then:
( a ) For any set of points $v_1,v_2,…v_k$, the union $U(v_1)∪U(v_2)∪…∪U(v_k)$ of their balls is not an n-disk.
( b ) For any two adjacent points u and v, there is a 1-sphere S(4) consisting of four points and containing points u and v (fig. 5.1).
( c ) For any two non-adjacent points u and v, their joint rim O(u)∩O(v) is not an (n-1)-disk.
Proof.
Assertion (a) follows from definition 5.1.
To prove ( b ), suppose that points v and u are adjacent (fig. 5.1). Suppose that the union C=U(v)∪U(u) is not an n-disk. Since O(v) and O(u) are (n-1)-spheres, then subspaces A=O(v)-u and B=O(u)-v are (n-1)-disks by theorem 4.1. Suppose that there is no connection between points belonging to IntA and IntB. Then A#B is an (n-1)-sphere and C is an n-disk by definition 4.4. This contains a contradiction because C is not an n-disk. Therefore, there are points w∈IntA and a∈IntB, which are adjacent. Hence, {v,u,a,w} is the 1-sphere.
To prove (c), suppose that u and v are non-adjacent points such that the intersection O(u)∩O(v) of their rims is an (n-1)-disk. Then the subspace U(v)+u containing point u and all points belonging to U(v) is an n-disk. From this contradiction, we conclude that O(u)∩O(v) is not an (n-1)-disk. □

Lemma 5.3.
Let N be a compressed n-manifold. If N is an n-sphere, then N is necessarily the minimal n-sphere $S_{min}$.
Proof.
Since N is an n-sphere, then N-v is an n-disk, where v is a point belonging to N, according to lemma 4.2. Since N is compressed, then N-v is the rim of some point u belonging to N, N-v=O(u). Therefore, O(v)=O(u) and N=$S^0$(v,u)⊕O(v) is the join of a 0-sphere $S^0$(v,u) and an (n-1)-sphere O(v). Take any point $v_1$ belonging to O(v). For the same reason as above, N=$S^0$($v_1$,$u_1$)⊕O($v_1$)=$S^0$($v_1$,$u_1$)⊕$S^0$(v,u)⊕O(v,$v_1$), where O(v,$v_1$) is the joint rim of adjacent points v and $v_1$ ((n-2)-sphere). Acting in the same way we finally obtain that N=$S^0$($v_n$,$u_n$)⊕⊕$S^0$($v_{n-1}$,$u_{n-1}$)⊕…$S^0$($v_1$,$u_1$)⊕$S^0$(v,u). □

Lemma 5.4.
( a ) Let N be a compressed n-manifold. If there are adjacent points v and u such that their rims O(v) and O(u) are minimal (n-1)-spheres, then N is the minimal n-sphere.
( b ) Let N be a compressed n-manifold. If for any two adjacent points v and u, their joint rim O(vu) is the minimal (n-2)-sphere, then N is the minimal n-sphere.
Proof.
To prove ( a ), suppose points v and u are adjacent and O(v) and O(u) are minimal (n-1)-spheres $S_1$ and $S_2$. Then O(v,u) is the minimal (n-2)-sphere S. Therefore, O(v)=$S^0$(au)⊕S and O(u)=$S^0$(b,v)⊕S. Since U(v)∪U(u) is not an n-disk, then by lemma 5.1, there is a 1-sphere S(4) containing points v and u and consisting of four points. Therefore, S(4) is necessarily {v,u,a,b}, where points a and b are adjacent. Hence, A=U(v)∪U(u)=$S^0$(a,u)⊕O(u)=$S^0$(a,u)⊕$S^0$(b,v)⊕S is the minimal n-sphere. Since A⊆N, then A=N by proposition 3.3. Hence, N is the minimal n-sphere. Assertion ( b ) can be proven similarly. □

The process of compression of an n-manifold by a sequence of d-transformations (which can be applied in some orders) can give a family of compressed spaces $G_1$,$G_2$,…$G_k$ which are not isomorphic to each other. However, if N is an n-sphere, then the process of compression always converts N to the minimal n-sphere.

6. Locally centered and lump covers of continuous closed n-manifolds and its properties.

In order to make this section self-contained we will use the necessary information from paper [11].
Suppose that a map h is a homeomorphism from $R^n$ to itself. If a set D is homeomorphic to a closed n-dimensional ball $B^n$ on $R^n$, then D is called a closed n-disk.
If a set S is homeomorphic to an n-dimensional sphere $S^n$ on $R^{n+1}$, then S is called an n-sphere.
We denote the interior and the boundary of an n-disk D by IntD and ∂D respectively. Since in this paper we use only closed n-disks, we say n-disk to abbreviate closed n-disk if no confusion can result.
Remind that collections of sets W={$D_1$,$D_2$,…} and U={$C_1$,$C_2$,…} are isomorphic (homotopic) if the intersection graphs G(W) and G(U) of W and U are isomorphic

(homotopic).

**Definition** 6.1.
Let collection $W=\{D_1,D_2,\ldots D_s\}$ of n-disks (in $R^m$, $m\geq n$). W is called a lump collection if:
- $D_1\cap D_2\cap\ldots D_s\neq\emptyset$.
- The intersection of any k distinct disks is an (n-k+1)-disk, $D_{i(1)}\cap D_{i(2)}\cap\ldots D_{i(k)}=D^{n-k+1}$.

(Fig. 6.1).

Facts about n-disks and spheres that we will need in this paper are stated below.

Fact 6.1.
If $D_1$ and $D_2$ are n-disks such that $D_1\cap D_2=D^{n-1}$ is an (n-1)-disk, then $D_1\cup D_2=B$ is an n-disk.

Fact 6.2.
Suppose that collection $W=\{D_1,D_2,\ldots D_s\}$ of n-disks (in $R^m$, $m\geq n$) is a lump collection. If a point v belongs to the interior of $D^{n-k+1}=D_1\cap D_2\cap\ldots D_k$, $v\in IntD^{n-k+1}$, then there is an open neighborhood O(v) of v belonging to the union $A=D_1\cup D_2\cup\ldots D_k$, $O(x)\subseteq A$, and homeomorphic to an open Euclidean n-ball.

Fact 6.3.
If S is an n-sphere and D is an n-disk belonging to S, then S-IntD is an n-disk.

Proposition 6.1.
Let collection $W=\{D_1,D_2,\ldots D_s\}$ of n-disks be a lump collection.
Then:
( a ) $s\leq n+1$.
( b ) The lump collection concept is hereditary: any subcollection of a lump collection is a lump collection itself.
( c ) Let $V=\{H_1,H_2,\ldots H_r\}$ be a collection of n-disks such that any $H_k=D_{k1}\cup D_{k2}\cup\ldots$ is the union of n-disks belonging to W and if $D_i\subseteq H_k$, then $D_i\not\subset H_p$, $p\neq k$. Then V is the lump collection of n-disks.
( d ) Collection $U=\{C_{p+1},C_{p+2},\ldots C_s\}$ where $C_{p+i}=D_1\cap D_2\cap\ldots D_p\cap D_{p+i}$ i=1,2,3,…s-p, is a lump collection of (n-p)-disks.
Proof.
The proofs of these assertions are simple so let us prove only proposition 6.1(c).
For dimensions n=1,2, it is verified directly. Assume that the proposition 6.1(c) is valid whenever n<a+1. Let n=a+1. With no loss of generality, consider collection $U=\{H_1,D_3,\ldots D_s\}$, where $D_1\cup D_2=H_1$. Obviously, $H_1$ is an n-disk and $H_1\cap D_3\cap\ldots D_s\neq\emptyset$. Then
$H_1\cap D_3\cap\ldots D_s=A\cup B$ where $A=D_1\cap D_3\cap\ldots D_s$ and $B=D_2\cap D_3\cap\ldots D_s$ are (n-s+2)-disks.
Since $A\cap B=D_1\cap D_2\cap D_3\cap\ldots D_s=C$ is an (n-s+1)-disk, then {A,B} is a lump collection. Hence, $A\cup B=H_1\cap D_3\cap\ldots D_s$ is an (n-s+2)-disk by the assumption.
Therefore, $U=\{H_1,D_3,\ldots D_s\}$ is a lump collection by definition 6.1. □

Corollary 6.1.
The union of any number of n-disks belonging to W is an n-disk, $D_{i1}\cup D_{i2}\cup\ldots D_{ip}=D$.

Helly's theorem [24] states that if a collection of convex sets in $E^n$ has the property that every (n + 1) members of the collection have nonempty intersection, then every finite subcollection of those convex sets has nonempty intersection. In application to digital modeling, this concept was studied in a number of works. In paper [23], a collection of convex n-polytopes possessing this property was called strongly normal (SN). One of the results was that if SN holds for every n+1 or fewer n-polytopes in a set of n-polytopes in $R^n$, then the entire set of n-polytopes is SN. In paper [9], a collection of sets with a similar property was called continuous. It was shown that the continuity of covers is necessary for digital models to be homotopic. In classical topology, the collection of sets W is centered if every finite subcollection of W has a point in common. This definition implies an infinite collection of sets. In this paper, we use only finite collections of sets. Since the word "normal" has already been used in the definition of a normal digital space [9], we define a locally centered collection (LCL-collection) as follows.

**Definition** 6.2.
Collection $W=\{D_1, D_2, \ldots D_s\}$ of n-disks is called locally centered if from condition $D_{i(k)} \cap D_{i(m)} \neq \varnothing$, k,m=1,2,..p, it follows that $D_{i(1)} \cap D_{i(2)} \cap \ldots D_{i(p)} \neq \varnothing$ (fig. 6.2)
Obviously, a lump collection is locally centered. The following proposition is an easy consequence of definition 6.2.

**Proposition** 6.2.
Let collection $W=\{D_1, D_2, \ldots D_s\}$ of n-disks be locally centered. Then:
- Any subcollection of W is locally centered.
- If $C_k=D_1 \cap D_k \neq \varnothing$, k=2,3,…s, then collection $U=\{C_2, C_3, \ldots C_s\}$ is locally centered and collections $V=\{D_2, D_3, \ldots D_s\}$ and $U=\{C_2, C_3, \ldots C_s\}$ are isomorphic.

Definition 6.3.
Let $W=\{D_1, D_2, \ldots D_s\}$ be a locally centered collection of n-disks such that if $D_{i(1)} \cap D_{i(2)} \cap \ldots D_{i(p)} \neq \varnothing$, then subcollection $V=\{D_{i(1)}, D_{i(2)}, \ldots D_{i(p)}\}$ is the lump one. Then W is called a locally centered lump collection (LCL collection) (fig. 6.2).

**Proposition** 6.3.
1. Let $W=\{D_0, D_1, \ldots D_s\}$ be an LCL collection of n-disks. Then any subcollection of W is an LCL collection of n-disks.
2. Let $W=\{D_1, D_2, \ldots D_s\}$ be an LCL collection of n-disks such that $D_1 \cap D_2 \cap \ldots D_k \cap D_m \neq \varnothing$, m=k+1,k+2,…s. Then collection $U=\{E_1, E_2, \ldots E_{s-k}\}$ where $E_i = D_1 \cap D_2 \cap \ldots D_k \cap D_{k+i}$, i=1,2,...s-k, is an LCL collection of (n-k)-disks belonging to the boundary of an (n-k+1)-disk $C=D_1 \cap D_2 \cap \ldots D_k$ and collections $X=\{D_{k+1}, D_{k+2}, \ldots D_s\}$ and $U=\{E_1, E_2, \ldots E_{s-k}\}$ are isomorphic.
3. Let $W=\{D_0, D_1, \ldots D_s\}$ be an LCL collection of n-disks such that $D_0 \cap D_i \neq \varnothing$, for i≠0. Then collection $V=\{C_1, D_2, \ldots D_s\}$ where $C_1 = D_0 \cup D_1$ is an LCL collection of n-disks such that $C_1 \cap D_i \neq \varnothing$ for i=2,3,…s and the union $C_1 \cup D_2 \cup \ldots D_s$ is an n-disk. ($D_1$ can be replaced by any $D_i$).
Proof.
Assertions 1 and 2 are checked directly. Let us prove assertion 3. To prove that collection V is locally centered, suppose that $C_1 \cap D_i \neq \varnothing$, $D_k \cap D_i \neq \varnothing$, i,k=2,3,…m. Since $D_0 \cap D_i \neq \varnothing$, i=2,3,…m, then $D_0 \cap D_2 \cap D_3 \cap \ldots D_m \neq \varnothing$ by definition 6.2 and $C_1 \cap D_2 \cap D_3 \cap \ldots D_m = (D_0 \cap D_2 \cap \ldots D_m) \cup (D_1 \cap D_2 \cap \ldots D_m) \neq \varnothing$. Therefore, V is locally centered.

Let $C_1 \cap D_2 \cap D_3 \cap \ldots D_m \neq \emptyset$. To prove that subcollection $Y=\{C_1,D_2,D_3,\ldots D_m\}$ is the lump one, it is enough to show that $C_1 \cap D_2 \cap D_3 \cap \ldots D_m = D$ is an (n-m+1)-disk. Obviously, $C_1 \cap D_2 \cap D_3 \cap \ldots D_m = (D_0 \cap D_2 \cap D_3 \cap \ldots D_m) \cup (D_1 \cap D_2 \cap D_3 \cap \ldots D_m)$. Suppose that $A = D_1 \cap D_2 \cap D_3 \cap \ldots D_m = \emptyset$. Since $D_0 \cap D_2 \cap D_3 \cap \ldots D_m \neq \emptyset$, then $B = D_0 \cap D_2 \cap D_3 \cap \ldots D_m$ is an (n-m+1)-disk by definitions 6.1 and 6.3. Therefore, $D=B$ is an (n-m+1)-disk. Suppose that $A = D_1 \cap D_2 \cap D_3 \cap \ldots D_m \neq \emptyset$. Then A is an (n-m+1)-disk and $A \cap B = D_0 \cap D_1 \cap D_2 \cap \ldots D_m = E$ is an (n-m)-disk by definitions 6.1 and 6.3. Therefore, $D = A \cup B$ is an (n-m+1)-disk by fact 6.1. Hence, $Y=\{C_1,D_2,D_3,\ldots D_m\}$ is a lump collection. □

We have already mentioned that in this paper, we consider continuous n-manifolds, which can be covered by a finite LCL collection of n-disks. By a continuous n-manifold, we mean a continuous closed (compact and without boundary) path-connected n-manifold. Since for a compact, each of its open covers has a finite subcover, then for a closed continuous n-manifold there is an LCL cover of it i.e., a closed continuous n-manifold can be decomposed as the union of n-disks belonging to its LCL cover.

**Proposition** 6.4.
Let an LCL collection $W=\{D_1,D_2\ldots D_t\}$ of n-disks be a cover of a closed n-manifold M.
( a ) Suppose $V=\{D_2,D_3\ldots D_p\}$ is the collection of all n-disks intersecting $D_1$ and $U=\{E_2,E_3,\ldots E_p\}$ is an LCL collection of (n-1)-disks such that $E_i = D_1 \cap D_i$, i=2,3,…p. Then U is an LCL cover of the boundary $\partial D_1$ of $D_1$, collections U and V are isomorphic and $C = D_1 \cup D_2 \cup \ldots D_p$ is an n-disk.
(b ) For any $D_i$ there exists $D_k$ such that $D_i \cap D_k = \emptyset$.
(c ) For any $D_i$ and $D_k$ such that $D_i \cap D_k \neq \emptyset$ there exist $D_p$ such that $D_i \cap D_p = \emptyset$, $D_k \cap D_p \neq \emptyset$.
(d ) $t \geq 2n+2$.
(Proof: see appendix.)

Remark 6.1.
For any n>0 there is an LCL cover $W=\{D_1,D_2\ldots D_t\}$ of n-sphere S by n-disks such that t=2n+2. Let us give an example of such a cover. Let $U^{n+1}$ be an (n+1)-dimensional cube in the Euclidean (n+1)-dimensional space $R^{n+1}$. Obviously, n-dimensional faces $F^n_k$, k=1,2,…2n+2, of $U^{n+1}$ form an LCL collection $W=\{F^n_1,F^n_2\ldots F^n_{2n+2}\}$ of n-dimensional disks. W is the minimal cover of an n-dimensional sphere, which is the boundary $\partial U^{n+1}$ of $U^{n+1}$.

LCL covers of a 1- and 2-spheres are depicted in fig. 6.3 and 6.4. The minimal LCL cover of a 1-sphere consists of four 1-disks, the minimal LCL cover of a 2-sphere consists of six 2-disks. Figure 6.5 shows examples of LCL tiling of a 2-plane and an LCL tessellation of Euclidean 3-space. An LCL cover of a 2-dimensional torus is depicted in fig 6.6.

Proposition 6.5.
Suppose that collection $W=\{D_0,D_1,D_2\ldots D_t\}$ of n-disks is an LCL cover of an n-sphere S and $V=\{D_1,D_2\ldots D_p\}$ is a subcollection of all n-disks intersecting $D_0$. Then collection U={$D_0,D_1,D_2…D_p,C$}, where C=$D_{p+1} \cup D_{p+2} \cup …D_t$, is an LCL cover of S by n-disks such that if $D_{i(1)} \cap D_{i(2)} \cap D_{i(m)} \neq \emptyset$, $D_{i(k)} \in V$, k=1,2,…m, then
$C \cap D_{i(1)} \cap D_{i(2)} \cap D_{i(m)} \neq \emptyset$.
(Proof: see appendix.)

Notice that the intersection graphs G(A) and G(B) of collections A={$D_0,D_1,D_2…D_p$} and B={$D_1,D_2…D_p,C$} are isomorphic.

Definition 6.4.
Suppose that collection W={$D_0,D_1,D_2…D_t$} of n-disks is an LCL cover of an n-sphere S and V={$D_1,D_2…D_p$} is a collection of all n-disks intersecting $D_0$. Then collection U={$D_1,D_2,…D_t$} is called a segmented n-disk, collection {$D_1,D_2…D_p$} is called the boundary $\partial U$ of U and collection {$D_{p+1},D_{p+2}…D_t$} is called the interior IntU of U (fig. 6.7).

Proposition 6.6.
Suppose that collection U={$D_1,D_2…D_t$} of n-disks is a segmented n-disk, collection $\partial U$={$D_1,D_2…D_p$} is the boundary of U and collection IntU={$D_{p+1},D_{p+2}…D_t$} is the interior of U. Then:
( a ) A=$D_1 \cup D_2… \cup D_t$ is an n-disk.
( b ) If $D_k \in$ IntU, then $\partial A \cap D_k = \emptyset$.
( c ) If $D_k \in \partial U$, then $\partial A \cap D_k = E_k$ is an (n-1)-disk and collection U={$E_1,E_2,…E_p$} is an LCL cover of the boundary $\partial A$ of A.
( d ) Collection V={$D_1,D_2…D_p,C$} where C=$D_{p+1} \cup D_{p+2}… \cup D_t$ is a segmented n-disk with only one interior n-disk C.
Proof.
Suppose that collection W={$D_0,D_1,D_2…D_t$} is an LCL cover of an n-sphere S according to definition 6.4.
To prove ( a ), notice that S=$D_0 \cup D_1… \cup D_t$. Then assertion ( a ) follows from fact 6.3.
To prove ( b ) and ( c ), notice that $\partial A = \partial D_0$. According to proposition 6.4, if $D_k \in$ IntU, then $D_0 \cap D_k = \emptyset$. If $D_k \in \partial U$, then $\partial A \cap D_k = D_0 \cap D_k = E_k$ is an (n-1)-disk and collection U={$E_1,E_2,…E_p$} is an LCL cover of the boundary $\partial A = \partial D_0$.
Assertion ( d ) follows from proposition 6.5. □

Definition 6.5.
Suppose that collection U={$D_1,D_2…D_t$} of n-disks is a segmented n-disk, collection $\partial U$={$D_1,D_2…D_p$} is the boundary of U and collection IntU={$D_{p+1},D_{p+2}…D_t$} is the interior of U. Denote by V={$D_1,D_2…D_p,C$} the LCL collection of n-disks, where C=$D_{p+1} \cup D_{p+2}… \cup D_t$. The replacing of collection U by collection V is called the merging of the interior of U or the merging of U. The replacing of collection V by collection U is called the splitting of the interior of V or the splitting of V. In this paper, this merging and splitting are called disk transformations or d-transformations of covers.
Obviously, A=$D_1 \cup D_2 \cup …D_t = D_1 \cup D_2 \cup …D_p \cup C$ is an n-disk.

The merging and the splitting of segmented n-disks belonging to LCL covers of a closed n-manifold M change the amount of elements in LCL covers of M. By the merging of segmented n-disks, we can reduce the amount of elements of a cover to the level where any segmented n-disk contains only one interior element. It is important that d-transformations do not change M itself. They change only the amount of elements in an LCL cover of M.

Definition 6.6.
Suppose that $W=\{D_0,D_1,D_2…D_t\}$ and $U=\{C_0,C_1,C_2…C_s\}$ are LCL covers of a closed n-manifold M. W and U are called equivalent if W can be converted into U by d-transformations.

Theorem 6.1.
( a ) Let an LCL collection $W=\{D_0,D_1,D_2…D_s\}$ of n-disks be a cover of an n-sphere S. Then any sequence of mergings necessarily converts W into the minimal LCL cover of S with 2n+2 elements.
( b ) Let an LCL collection $W=\{D_0,D_1,D_2…D_s\}$ of n-disks be a cover of a closed n-manifold M. Then a sequence of d-transformations of W converts W into another LCL cover of M.
Proof.
( a ) Suppose that collection $W=\{D_0,D_1,D_2…D_t\}$ of n-disks is an LCL cover of an n-sphere S and $V=\{D_1,D_2…D_p\}$ is the collection of n-disks intersecting $D_0$. Then collection $U=\{D_1,D_2…D_t\}$ is a segmented n-disk with the boundary $\partial U=\{D_1,D_2…D_p\}$ and the interior $IntU=\{D_{p+1},D_{p+2}…D_t\}$. Let us convert U to $V=\{D_1,D_2…D_p,C\}$ by replacing the collection of interior disks by $C=D_{p+1}\cup D_{p+2}…\cup D_t$. According to proposition 6.5, collection $W_1=\{D_0,D_1,D_2…D_p,C\}$ is an LCL cover of S by n-disks. Take element $D_1$ instead of $D_0$ and apply to $U_1=\{D_0,D_2…D_p,C\}$ the same operation. Finally, we obtain the minimal LCL cover of S consisting of 2n+2 elements according to remark 6.1.
( b ) To prove (b), suppose that $W=\{D_1,D_2…D_t\}$ is an LCL cover of a closed n-manifold M, $U=\{D_1,D_2…D_s\}$ is a segmented n-disk, $A=D_1\cup D_2…\cup D_s$ is an n-disk, $\partial U=\{D_1,D_2…D_m\}$ is The boundary of U (and subcollection of U containing all n-disks intersecting the boundary $\partial A$ of A) and $IntU=\{D_{m+1},D_{m+2}…D_s\}$ is the interior of U (and a subcollection of U containing all n-disks not intersecting the boundary $\partial A$). Merge all n-disks belonging to IntU to $C=D_{m+1}\cup D_{m+2}\cup…D_s$. Then collection $V=\{D_1,D_2,…D_m,C,D_{s+1}…D_t\}$ is a cover of M. Note that $V_1=\{D_1,D_2,…D_m,C\}$ is an LCL collection as it follows from proposition 6.5, $V_2=\{D_{s+1}…D_t\}$ is too an LCL collection according to proposition 6.3. Since $C\cap D_i=\varnothing$, i=s+1,s+2,…t, then V is an LCL collection. The proof is complete. □

An irreducible LCL cover of a 2-dimensional torus is illustrated in fig. 6.6.

Problem 6.1.
There is an open problem similar to problem 4.1. Suppose that $W=\{D_0,D_1,D_2…D_t\}$ and $U=\{C_0,C_1,C_2…C_s\}$ are LCL covers of the same closed n-manifold M. Can W be converted to U by a sequence of d-transformations?
The answer yes is only for a continuous n-sphere.

7. A connection between digital spaces and LCL covers of continuous n-manifolds.

For technical convenience, call the collection of sets $W=\{u_1,u_2,…\}$ contractible, if the intersection graph G(W) of W is contractible, homotopic if G(W) and G(U) are homotopic and isomorphic if G(W) and G(U) are isomorphic.

A part of this section with technical results leading to the proof of theorem 7.1 is placed in appendix

Theorem 7.1.
Let W={$D_1,…D_s$} be an LCL collection of n-disks and A=$D_1 \cup D_2 \cup …D_s$. The intersection graph G(W) of W is contractible if and only if A is an n-disk (fig. 7.1).

Lemma 7.1.
Let an LCL collection W={$D_0,D_1,D_2…D_s$} of n-disks be a cover of a closed n-manifold M. Then the intersection graph G(W) of W is a digital n-manifold.
Proof.
The proof is by induction. For n=1, the theorem is plainly true for s≥4 (fig. 6.3). Assume that the theorem is valid whenever n<a+1. Let n=a+1.
For definiteness, consider n-disk $D_0$, subcollection O($D_0$)={$D_1,D_2…D_r$} of all n-disks intersecting $D_0$ without $D_0$ and collection V={$C_1,C_2,…C_r$} where $C_i$=$D_0 \cap D_i$. By proposition 6.3, V is an LCL collection of (n-1)-disks and collections V and O($D_0$) are isomorphic.
Obviously, V is a cover of (n-1)-sphere $\partial D_0$ such that $\partial D_0$=$C_1 \cup C_2 \cup …C_r$. Therefore, the intersection graph G(V) of V is a digital (n-1)-manifold ((n-1)-sphere) by the induction hypothesis. Since collections V and O($D_0$) are isomorphic by proposition 6.3, the intersection graphs G( O($D_0$)) of O($D_0$) and G(V) of V are isomorphic. Therefore, G( O($D_0$)) is a digital (n-1)-manifold. Hence, the intersection graph G(W) of collection W is a digital n-manifold by definition 4.1. This completes the proof. □

Lemma 7.2.
Let W={$D_0,D_1,D_2…D_s$} be an LCL collection of n-disks. If the intersection graph G(W) of W is a digital n-manifold, then W is the LCL cover of a closed continuous n-manifold M=$D_0 \cup D_1 \cup …D_s$.
The proof is similar to the proof of lemma 7.1 and hence omitted.

The following theorem summarizes the results of lemmas 7.1 and 7.2.

Theorem 7.2.
Let W={$D_0,D_1…D_t$} be an LCL collection of n-disks. W is a cover of a closed n-manifold M=$D_1 \cup D_2 \cup …D_s$ if and only if the intersection graph G(W) of W is a digital n-manifold.

Lemma 7.3.
Let W={$D_1,D_2…D_s$} be an LCL collection of n-disks. If W is a segmented n-disk, then the intersection graph G(W) of W is a digital n-disk.
Proof.
Suppose that W is a segmented d n-disk. Then A=$D_1 \cup D_2…\cup D_s$ is an n-disk according to proposition 6.6 and the intersection graph G(W) of W is contractible according to theorem 7.1. According to definition 6.4, there is an LCL collection U={$D_0,D_1…D_s$} of n-disks such that S=$D_0 \cup D_1 \cup …D_s$ is an n-sphere. Suppose that G(U)={$v_0,v_1…v_s$} is the intersection graph of U where f($D_k$)=$v_k$, k=0,1,…s. According to theorem 7.2, G(U) is a digital n-manifold. Then G(W)=G(U)-$v_0$ is a digital n-manifold with boundary. Therefore, G(W) is a digital n-disk according to

definition 4.2. □

The following lemma is an easy consequence of lemma 7.2.
Lemma 7.4.
Let $W=\{D_1,D_2…D_s\}$ be an LCL collection of n-disks. If the intersection graph $G(W)$ of W is a digital n-disk, then W is a segmented n-disk.

Theorem 7.3 summarizes the results of lemma 7.3 and 7.4.

Theorem 7.3.
Let $W=\{D_1,D_2…D_s\}$ be an LCL collection of n-disks. W is a segmented n-disk if and only if the intersection graph $G(W)$ of W is a digital n-disk.

Corollary 7.1.
Let $W=\{D_1,D_2…D_s\}$ be an LCL collection of n-disks. W is a cover of a continuous n-sphere if and only if the intersection graph $G(W)$ of W is a digital n-sphere.

Let collection $W=\{D_1,D_2…D_s,…\}$ be an LCL tiling of n-dimensional Euclidean space $R^n$ by n-disks. Then for the collection $O(D_i)=\{D_{i(1)},D_{i(2)}…D_{i(k)}\}$ of all disks intersecting any given n-disk $D_i$ (without $D_i$), the intersection graph $G(O(D_i))$ of this collection is a digital (n-1)-sphere. Therefore, the intersection graph $G(W)=Y^n$ of W is a digital n-manifold. $Y^n$ is a digital model of continuous n-dimensional Euclidean space. $Y^n$ can be constructed in a variety of ways depending on the choice of tiling W. Figure 6.5 depicts LCL covers of continuous 2- and 3-dimensional Euclidean spaces.

Theorem 7.4.
Suppose that collection $W=\{D_0,D_1,D_2,…D_s\}$ of n-disks is an LCL cover of a closed n-manifold M and $G(W)$ is the intersection graph of W. Then:
( a ) Any d-transformation of W generates the d-transformation of $G(W)$.
( b ) Any d-transformation of $G(W)$ generates the d-transformation of collection W.
Proof.
1. To prove (a), suppose that $W=\{D_0,D_1,D_2,…D_t\}$ is an LCL cover of M, $V=\{D_0,D_1,D_2,…D_s\}$ is a segmented n-disk, $X=\{D_0,D_1,D_2,…D_m\}$ is the boundary of V and $Y=\{D_{m+1},D_{m+2},…D_s\}$ is the interior of V.
Then by theorems 7.2 and 7.3, $G(W)=\{v_0,v_1,v_2,…v_t\}$ is a digital closed n-manifold, $f(D_i)=v_i$, i=0,1,2,…t, $G(V)=\{v_0,v_1,v_2,…v_s\}$ is a digital n-disk, $G(X)=\{v_0,v_1,v_2,…v_m\}$ is the boundary of $G(V)$ and $G(Y)=\{v_{m+1},v_{m+2},…v_s\}$ is the interior of $G(V)$.
The merging of all n-disks belonging to Y into an n-disk $C=D_{m+1}\cup D_{m+2}\cup…D_s$ according to definition 6.5, generates the replacing of all points belonging to $G(Y)$ by one point u which is adjacent to all points belonging to $G(X)$. This replacing is the d-transformation of $G(W)$
2. (b) can be proved applying the same procedure in the reverse order. □

Lemma 7.5.
Suppose that closed continuous n-manifolds M and N are not homeomorphic. Then any two LCL covers $W=\{D_0,D_1,D_2,…D_s\}$ and $U=\{C_0,C_1,C_2,…C_t\}$ of M and N are not isomorphic.
Proof.
Assume that there is an LCL cover $U=\{C_0,C_1,C_2,…C_s\}$ of N such $G(U)=G(W)$. Then there can be established a homeomorphism between $D_0\cup D_1\cup D_2…\cup D_s$ and $C_0 \cup C_1 \cup C_2 \ldots \cup C_s$. Since $M = D_0 \cup D_1 \cup D_2 \ldots \cup D_s$ and $N = C_0 \cup C_1 \cup C_2 \ldots \cup C_s$ are not homeomorphic, then the assumption is not valid. □

Let us introduce a construction, which will be used in further proofs.

Definition 7.1.
Let $W = \{D_0, D_1, D_2, \ldots D_s\}$ be an LCL collection of n-disks. W is called a segmented k-disk, k-sphere or k-manifold, $k \leq n$, if the intersection graph G(W) of W is a digital k-disk, k-sphere or k-manifold respectively (fig. 7.2).
If W is a segmented k-disk, then the boundary $\partial W$ of W is a segmented (k-1)-sphere belonging to W and such that $G(\partial W)$ is the boundary of G(W). The interior IntW of W is defined as $W - \partial W$.
Obviously, if k=n, then a segmented n-disk by definition 7.1 is a segmented n-disk by definition 6.4.

Remark 7.1.
As in the digital approach, d-transformations of an LCL cover W of a closed n-manifold M can change segmented m-spheres and m-disks belonging to W, m<n.
Suppose that collection $W = \{D_0, D_1, D_2, \ldots D_s\}$ of n-disks is an LCL cover of a closed n-manifold M, S is a segmented m-sphere belonging to W, C is a segmented m-disk belonging to W, m<n. We say that S is embedded in D if there is a segmented n-disk D such that $S \subseteq D$. If $S \cap IntD \neq \emptyset$, then after the merging of all interior elements of D into an element w, S is collapsed into a set of elements which is not a segmented m-sphere. If $S \subseteq \partial D$, then after the merging of all interior elements of D into an element w, S belongs to the boundary of a segmented n-disk $D_1$ with only one interior element w (fig, 4.6).
We say that C is embedded in D if there is a segmented n-disk D such that $IntC \subseteq IntD$ (and $C \subseteq D$). If $\partial C \cap IntD \neq \emptyset$, then after the merging of all interior elements of D into an element w, D is collapsed into a set of elements, which is not a segmented m-disk. If $\partial C \subseteq \partial D$, then after the merging of all interior elements of D into an element w, C converts into a segmented m-disk $v \oplus \partial C$ with only one interior element w.
Obviously, a segmented m-sphere S (a segmented m-disk C) is embedded into a segmented n-disk D if and only if the intersection graph G(S) (G(C)) is embedded into the intersection graph G(D).
Further for technical convenience, n-disks belonging to an LCL collection W are called elements of W.

Remark 7.2.
Suppose that $W = \{w_1, w_2, \ldots w_t\}$ is an LCL cover of a closed 3-manifold M. For a segmented 1-sphere $S = \{w_1, w_2, \ldots w_p\}$ belonging to W there is a continuous closed curve C belonging to M with the following properties.
 $C \subseteq w_1 \cup w_2 \ldots \cup w_p = A$.
$w_k \cap C$, $k = 1, 2, \ldots p$ is a closed 1-disk.
We say that C is a continuous analog of S and S is a segmented analog of C (fig. 7.3).
Similarly, for a segmented 2-disk $U = \{w_1, w_2, \ldots w_p\}$ belonging to W there is a continuous 2-disk D belonging to M with the following properties.
$D \subseteq w_1 \cup w_2 \ldots \cup w_p = A$.
$w_k \cap D$, $k = 1, 2, \ldots p$ is a closed 2-disk.
We say that D is a continuous analog of U and U is a segmented analog of D (fig.

7.3).

8. Properties of a continuous 2-sphere.

Obviously, a digital normal 2-space is necessarily a digital 2-manifold.
Let us apply the previous results to 2-manifolds and consider conditions about the characterization of the continuous 2-sphere amongst continuous closed 2-manifolds using LCL covers and their intersection graphs, which are digital 2-manifolds. As it follows from the previous results, any LCL cover of a closed continuous n-manifold can be converted to an irreducible cover by the merging of elements of the cover. If a closed continuous n-manifold M is an n-sphere, then any LCL cover of M can be necessarily converted to the minimal cover, which is the collection $W=\{F_1,F_2\ldots F_{2n+2}\}$ of n-dimensional faces $F_k$, k=1,2,…2n+2, of an (n+1)-dimensional cube U.
Let us first prove a digital theorem whose results will be used in further proofs.

Theorem 8.1.
Let H be a digital 2-manifold. Suppose that for H and for any digital 2-manifold equivalent to H, any digital 1-sphere S belonging to H is the boundary of some digital 2-disk D belonging to H, then H is a digital 2-sphere.
Proof.
Let $S \subseteq H$ be a 1-sphere, D be a 2-disk such that $S=\partial D$ and v be a point not belonging to H. Replace D with the 2-disk $v \oplus \partial D$ by the deleting of all points belonging to IntD and connecting point v with any point of $\partial D$. This is a d-transformation, which converts D into $v \oplus S$. Repeat this procedure until any 1-sphere S is the rim of some point v and any 2-disk is the join of a point v and a 1-sphere S. Denote the obtained space by G.
Note that G is a compressed digital 2-manifold equivalent to H. Suppose $\{v_1,v_2,\ldots v_s\}$ are points of G. According to lemma 5.2, for any two adjacent points, say $v_1$ and $v_2$, there is a digital 1-sphere, say $S_1=\{v_1,v_2,v_3,v_4\}$, containing four points. Since any digital 1-sphere is the rim of some point, then there is a point, say $v_5$, adjacent to $v_1,v_2,v_3$ and $v_4$ (fig. 5.1, 8.1). For the same reason, adjacent points $v_1$ and $v_5$ belong to a digital 1-sphere $S_2$ consisting of four points and $S_2$ is the rim of some point $v_i$. Obviously, point $v_i$ is either $v_2$ or $v_4$. Let $v_2$ be $v_i$. Then $S_2=\{v_1,v_5,v_3,v_6\}$. Applying the above arguments to adjacent points $v_5$ and $v_4$, we see that points $v_4$ and $v_6$ must be adjacent. Obviously, subspace $G_1 \subseteq G$ containing point $\{v_1,v_2,\ldots v_6\}$ is the minimal digital 2-sphere, $G_1=S^2_{min}$. According to proposition 3.3, if $G_1 \subseteq G$, then $G_1=G$.
Therefore, H is a digital 2-sphere by theorem 4.1. The proof is complete. □

Theorem 8.2.
Let M be a closed continuous 2-manifold. If for any LCL cover W of M by 2-disks, any segmented 1-sphere belonging to W is the boundary of some segmented 2-disk belonging to W, then M is a continuous 2-sphere.
Proof.
Suppose that $W=\{w_1,w_2,\ldots w_t\}$ is an LCL cover of M by 2-disks and S is a segmented 1-sphere (fig. 8.1). According to theorems 7.2 and 7.4, the intersection graph G(W) is a digital 2-manifold. Since for any segmented 1-sphere S (fig. 8.1) belonging to W there is a segmented 2-disk $D=\{w_1,w_2,\ldots w_p\}$ belonging to W such that $S \subseteq D$, then for any digital 1-sphere X=G(S) belonging to G(W) there is a digital 2-disk Y=G(D) belonging to G(W) such that $X \subseteq Y$. Therefore, G(W) satisfies to conditions of theorem 8.1. By theorem 8.1, G(W) is equivalent to the minimal 2-sphere $S^2_{min}$ with points $\{v_1, v_2, \ldots v_6\}$. According to theorems 7.2 and 7.4, there is an LCL cover $U=\{u_1, u_2, \ldots u_6\}$ of M equivalent to W and such that the intersection graph G(U) of U is $S^2_{min}$. Obviously, $S^2_{min}$ is the intersection graph G(F) of collection $F=\{F_1, F_2, \ldots F_6\}$ of 2-faces $F_k$, k=1,2,…6, of a continuous 3-cube U (fig. 8.1). There is an obvious homeomorphism between $M = u_1 \cup u_2 \cup \ldots u_6$ and the boundary $\partial U = F_1 \cup F_2 \cup \ldots F_6$ of U. Hence, M is a continuous 2-sphere. The proof is complete. □

We can change the conditions of theorem 8.1 and 8.2 as follows.

Theorem 8.3.
Let H be a digital 2-manifold. If for H, for any digital 2-manifold equivalent to H and for any digital 1-disk L belonging to H there is a digital 2-disk D belonging to H and such that IntL⊆IntD, then H is a digital 2-sphere.
Proof.
As in the proof of theorem 8.1, convert H into a compressed 2-manifold G by d-transformations according lemma 5.1.
For a point v∈G, take some point u belonging to a 1-sphere O(v) (fig. 8.2). Then $O(v,u) = O(v) \cap O(u) = S(x,y)$ is a 0-sphere by proposition 3.2 and O(v)-u is a 1-disk L by lemma 4.2. For any 1-disk there is a digital 2-disk D belonging to G and such that IntL⊆IntD. Since any 2-disk is the ball of some point, then $D=U(a_1)$ and L consists of three points x, y and $a_1$.
Therefore, $O(v) = S_{min} = S(x,y) \oplus S(u,a_1)$. Applying the above arguments to any other point in G, we see that the rim of any point is $S_{min}$ consisting of four points.
According to lemma 4.4, G is the minimal 2-sphere. Therefore, H is a digital 2-sphere. □

Theorem 8.4.
Let M be a closed continuous 2-manifold. If for any LCL cover W of M by 2-disks and for any segmented 1-disk L belonging to W there is a segmented 2-disk D belonging to W and such that IntL⊆IntD, then M is a continuous 2-sphere.
The proof is similar to the proof of theorem 8.2 and is omitted.

It can be checked directly for a digital 2-dimensional projective plane P that there are digital 1-disks belonging to P which can not be contracted to a point (fig. 4.5). An LCL cover of a 2-dimensional continuous torus is depicted in fig. 6.6. Any segmented 1-disks containing 5 elements of the cover can not be contracted to a point.

9. Properties of a continuous 3-sphere.

At first, let us briefly review some topological notions related to the Poincaré conjecture.

Here is the standard form of the Poincaré conjecture:
Every simply connected closed 3-manifold is homeomorphic to a 3-sphere.

A closed 3-manifold M is called simply connected if and only if M is path-connected and the fundamental group of M is trivial, i.e. consists only of the identity element.

Loosely speaking, if the fundamental group of M is trivial, then any closed curve belonging to M can be continuously shrunken to a point.

Our approach is based on a decomposition of a closed 3-manifold M into an LCL collection of 3-disks with certain properties. We study an LCL cover of a manifold instead of the study of the manifold itself. We do not use a mapping from a circle or a 2-disk into a closed 3-manifold. We introduce segmented 1-, 2- and 3-disks and spheres (according to definition 7.1) which are segmented analogs of continuous 1-, 2- and 3-disks and spheres belonging to a closed 3-manifold. We have to find conditions, which guarantee that a closed continuous 3-manifold is a continuous 3-sphere.

It is important to emphasize that d-transformations of an LCL cover of M do not change the manifold itself. They change only a cover of M by converting one LCL cover into another LCL cover.

Let us first prove a digital theorem whose results will be used in the further proof.

Theorem 9.1.
Let H be a digital 3-manifold and W(H) be a collection of digital 3-manifolds equivalent to H. If for any digital 2-disk D belonging to any H∈W(H) there is a digital 3-disk U belonging to H∈W(H) such that IntD⊆IntU, then H is the digital 3-sphere.

 Proof.
Let U be a digital 3-disk belonging to H. Suppose that a point v belongs IntU. Delete all points belonging to IntU except point v and connect point v with all points belonging to ∂U. This is a d-transformation that converts U into $U_1$=v⊕∂U. Repeat this procedure until any digital 3-disk U is the ball of some point according to lemma 5.1.

Denote by G the obtained digital 3-manifold. Clearly, G is equivalent to H because all replacings are d-transformations.

For a point v∈G, take some point u belonging to the rim O(v) of v. Note that O(v) is a digital 2-sphere. (fig. 9.1). Then O(v,u)=S is a digital 1-sphere and O(v)-u is a digital 2-disk D according to lemma 4.2. Therefore, there is some digital 3-sisk U such that IntD belongs to IntU. Since any digital 3-disk is the ball of a point of G, then there is a point $u_2$ such IntU=$u_2$. Since IntD⊆IntU($u_2$), then IntD=$u_2$, v∈O($u_2$) and O(v,$u_2$)=S. Therefore, O(v)= S(u,$u_2$)⊕S.

Take a point $v_1$ belonging to O(v,u)=S and apply to $v_1$ the same arguments as above. We obtain that O(v)=S(u,$u_2$)⊕S($v_1$,$v_2$)⊕S($w_1$,$w_2$). Therefore, O(v) is the minimal digital 2-sphere $S^2_{min}$. Since point v is chosen arbitrarily, then the rim of any point in G is $S^2_{min}$.

Then according to lemma 4.4, G is the minimal digital 3-sphere $S^3_{min}$ with eight points {u,$u_2$,$v_1$,$v_2$,$w_1$,$w_2$,v, p}. (fig. 9.1). Hence, H is a digital 3-sphere according to theorem 4.1. The proof is complete. □

In this theorem, a digital loop is presented only implicitly, as the boundary of a digital 2-disk.

Theorem 9.2.
Let M be a closed continuous 3-manifold. If for any LCL cover W of M by 3-disks and for any segmented 2-disk D belonging to W there is a segmented 3-disk U belonging to W such that IntD⊆IntU, then M is a continuous 3-sphere.
Proof.
Suppose that W={$w_1,w_2,...w_t$} is an LCL cover of M by 3-disks and G(W) with points {$v_1,v_2,...v_t$} is the intersection graph of W, where $w_i$ corresponds $v_i$. According to theorem 7.2, G(W) is a digital 3-manifold satisfying to conditions of theorem 9.1. Therefore, G(W) can be transformed to the digital minimal 3-sphere $S^3_{min}$ with points {$v_1,v_2,...v_8$} by d-transformations. According to theorem 7.4, W can be transformed to an LCL cover $W_1$={$b_1,b_2,...b_8$} consisting of eight elements and such that the intersection graph $G(W_1)$ of $W_1$ is $S^3_{min}$. According to proposition 6.4 and remark 6.1, the minimal digital 3-sphere $S^3_{min}$ is the intersection graph G(F) of collection F={$F_1,F_2,...F_8$} of 3-dimensional faces of the unit continuous four-dimensional cube $U^4$. There is an obvious homeomorphism between M=$b_1\cup b_2\cup...b_8$ and $\partial U^4$=$F_1\cup F_2\cup...F_8$. Therefore, M is a continuous 3-dimensional sphere. The proof is complete. □

A geometrical sense of this theorem is intuitively clear.
Suppose that M is a closed path-connected (continuous) 3-manifold, W={$w_1,w_2,...w_t$} is an LCL cover of M by 3-disks.
Suppose that M is a continuous 3-sphere and D is a closed 2-disk belonging to M. If there is a segmented 2-disk U containing D, then there is a segmented 3-disk V containing U and D.
Suppose that M is not homeomorphic to a 3-sphere. Then there exists a segmented 2-disk U such that there is no segmented 3-disk containing U. Therefore, for a continuous 2-disk D belonging to M and such that D is a continuous analog of U, there is no segmented 3-disk containing D. U and D are just too large for being contained in a segmented 3-disk and there are not enough elements left in W in order to form a segmented 3-disk containing U and D. Therefore, there is a continuous closed curve C (for example, the boundary of D, C=$\partial$D) such that its segmented analog - a segmented 1-sphere does not belong to any segmented 3-disk belonging to W.
This property resembles the condition used by Bing who showed that a simply-connected, closed 3-manifold with the property that every loop is contained in a 3-ball is homeomorphic to the 3-sphere.
As it is seen from theorem 9.2, we do not use segmented loops explicitly and therefore, we do not need to impose any restrictions or requirements on them. Implicitly, a segmented closed curve is presented only as the boundary of a segmented 2-disk.

10. A connection between the classification problem for closed continuous 3-manifolds and digital 3-manifolds.

Possibly, the approach presented in this paper can help in treating the problem of classification of compact 3-dimensional manifolds. The advantage of this approach is that a continuous closed 3-manifold can be presented as a digital 3-manifold and, therefore, investigated by means of computers.
Suppose that W={$w_1,w_2,...w_t$} is an LCL cover of a closed 3-manifold M. By the merging or splitting of 3-disks belonging to W, the amount of elements of W can be reduced or increased. According to theorems 7.2 and 7.4, the intersection graph G(W) of W is a digital 3-manifold and d-transformations of W generate d-transformations of G(W). Conversely if G(W) is a digital n-manifold, then an LCL collection is a cover of a closed (continuous) 3-manifold.
Therefore, if we can classify digital 3-manifolds, then this classification can be applied to continuous closed 3-manifolds.
On the first step, digital 3-manifolds can be distinguished by an amount of points contained in their compressed versions. Obviously, for any digital 3-manifold there always exist one or several compressed versions with an equally small amount of points. Suppose that E(H) is a family of all digital 3-manifolds equivalent to a digital n-manifold H and $\{G_1, G_2, \ldots G_k\}$ is a family of 3-manifolds belonging to E(H) with the minimal amount p of points among manifolds belonging to E(H). Let us denote p as the class number p(E(H)) or p(H).
Therefore, for any digital 3-manifold H there is a unique p(H). If H and F are digital 3-manifolds such that p(H)≠p(F), then H and F are not equivalent.

The following table shows the class number p(H) for some digital manifolds.

| Manifold H | $S^0$ | $S^1$ | $S^2$ | $P^2$ | $T^2$ | $S^3$ | $S^4$ | $S^5$ | $S^6$ |
|---|---|---|---|---|---|---|---|---|---|
| p(H) | 2 | 4 | 6 | 11 | 16 | 8 | 10 | 12 | 14 |

Here $S^n$ is a digital n-sphere, $P^2$ is a digital 2-dimensional projective plane, $T^2$ is a digital 2-dimensional torus.
It is clear that for some integers q there is no H such that q=p(H). For example, for q=10 there is no digital n-manifold H such that p(H)=10.

Suppose that M is a continuous closed 3-manifold, W is an LCL cover of M by 3-disks and E(W) is a collection of LCL covers of M equivalent to W. If H=G(W) is the intersection graph of W, then p(H) is the unique number characterizing M and E(W).

Classification on the second level can be based on properties of digital 3-manifolds without a point.

Lemma 10.1.
Suppose that H is a digital n-manifold and points u and v belong to H. Then a digital n-manifold with boundary H-u is homotopic to a digital manifold with boundary H-v.
Proof.
Since a digital manifold is connected, then there is a path $\{u=v_1, v_2, \ldots v_k=v\}$ where $v_k$ is adjacent to $v_{k-1}$ and $v_{k+1}$ and non-adjacent with all other points belonging to the path. Delete point $v_1$ from H. Then the rim $A(v_2)$ of point $v_2$ in H-$v_1$ is $O(v_2)$-$v_1$ where $O(v_2)$ is the rim of $v_2$ in H. Since $O(v_2)$ is a digital (n-1)-sphere, then $O(v_2)$-$v_1$ is a digital (n-1)-disk. Therefore, $O(v_2)$-$v_1$ is contractible and point $v_2$ can be deleted from H-$v_1$ according to definition 3.2. In H-$v_1$-$v_2$, subspace $O(v_1)$-$v_2$ is contractible ((n-1)-disk). Therefore, point $v_1$ can be glued to H-$v_1$-$v_2$ in such a manner that $v_1$ is connected to all points in $O(v_1)$-$v_2$. The obtained space is H-$v_2$. Therefore, H-$v_2$ is homotopic to H-$v_1$. Similarly, H-$v_i$ is homotopic to H-$v_{i+1}$, i=1,2,…k-1. Hence, H-u is homotopic to H-v. □

Lemma 10.2.
Suppose that digital n-manifolds H and F are equivalent and points u and v belong to H and F respectively. Then H-u and F-v are homotopic.
Proof.
Suppose that $f_i$, i=1,2,…k is a d-transformation and F=$f_k \ldots f_2 f_1$H. A d-transformation

$f_1$ involves points of a digital n-disk D belonging to H. Obviously, there is a point $u_1$ belonging to H and such that the rim $O(u_1)$ of $u_1$ does not intersect IntD. Therefore, $f_1$ is a d-transformation applied to $H-u_1$ and $f_1(H-u_1)=(f_1H)-u_1$ is equivalent to $H-u_1$. By lemma 10.2, $H-u_1$ is homotopic to H-u. Hence, $f_1H-u_1$ is homotopic to H-u. For $H_1=f_1H$, there is a point $u_2$ such that $O(u_2)$ is not involved in a d-transformation $f_2$. For the same reason as above, $f_1H_1-u_2$ is homotopic to H-u. Finally, we obtain that F-v is homotopic to H-u. □

This lemma means that if H-u and F-v are not homotopic, then H and F are not equivalent.
As an illustration, consider n-manifolds without a point (see fig 4.4 for $S^2$-v and $P^2$-v and fig.6.6 for $T^2$).

| manifold | $S^1$-v | $S^2$-v | $S^3$-v | $P^2$-v | $T^2$-v |
|---|---|---|---|---|---|
| Homotopic space |  |  |  | 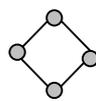 | 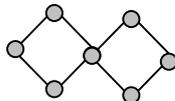 |

Here are some of the results established in this paper.
• Let H be a digital n-manifold. H is a digital n-sphere if and only if for any point v belonging to H, H-v is a digital n-disk.
• Let H be a digital n-manifold. H is a digital n-sphere if and only if H is equivalent to the minimal digital n-sphere.
• W is an LCL cover of a continuous closed n-manifold M if and only if the intersection graph G(W) of W is a digital n-manifold.
• Let M be a closed continuous 2-manifold. If for any LCL cover W of M by 2-disks, any segmented 1-sphere belonging to W is the boundary of some segmented 2-disk belonging to W, then M is a continuous 2-sphere.
• Let M be a closed continuous 2-manifold. If for any LCL cover W of M by 2-disks and for any segmented 1-disk L belonging to W there is a segmented 2-disk D belonging to W and such that IntL⊆IntD, then M is a continuous 2-sphere.
• Let M be a closed continuous 3-manifold. If for any LCL cover W of M by 3-disks and for any segmented 2-disk D belonging to W there is a segmented 3-disk U belonging to W such that IntD⊆IntU, then M is a continuous 3-sphere.

References.

Appendix.

Proposition 6.4.
Let an LCL collection $W=\{D_1,D_2\ldots D_t\}$ of n-disks be a cover of a closed n-manifold M.
( a ) Suppose $V=\{D_2,D_3\ldots D_p\}$ is the collection of all n-disks intersecting $D_1$ and $U=\{E_2,E_3,\ldots E_p\}$ is an LCL collection of (n-1)-disks such that $E_i=D_0\cap D_i$, $i=2,3,\ldots p$. Then U is an LCL cover of the boundary $\partial D_1$ of $D_1$, collections U and V are isomorphic and $C=D_1\cup D_2\cup\ldots D_p$ is an n-disk.
(b ) For any $D_i$ there exists $D_k$ such that $D_i\cap D_k=\varnothing$.
(c ) For any $D_i$ and $D_k$ such that $D_i\cap D_k\neq\varnothing$ there exist $D_p$ such that $D_i\cap D_p=\varnothing$, $D_k\cap D_p\neq\varnothing$.
(d ) $t\geq 2n+2$.
Proof.
Assertion ( a ) follows directly from proposition 6.3.
To prove ( b ) and ( c ), suppose that subcollection $V=\{D_1,D_2\ldots D_k\}$ contains all n-disks intersecting $D_1$ including $D_1$. Then the union $C_k=D_1\cup D_2\cup\ldots D_k$ is an n-dimensional disk by proposition 6.3. Therefore, V is not a cover of M and there is at least one n-disk, which does not intersect $D_1$.
Suppose that $U=\{D_1,D_3,D_4\ldots D_m,D_{p+1},D_{p+2}\ldots D_{p+h}\}$ is the collection of all n-disks belonging to W and intersecting $D_2$, where $D_i\in V$, $i=3,4,\ldots m$, $D_i\notin V$, $i=p+1,p+2,\ldots p+h$. Then $X=\{E_1,E_3,E_4\ldots E_m,E_{p+1},E_{p+2}\ldots E_{p+h}\}$ where $E_i=D_2\cap D_i$, $D_i\in U$, is an LCL cover of an (n-1)-sphere $\partial D_2$ by (n-1)-disks according to corollary 6.1. According to proposition 6.3, $E_1\cup E_3\cup\ldots E_m$ is an (n-1)-disk. Therefore, h>0 and at least $E_{p+1}=D_2\cap D_{p+1}\neq\varnothing$. Therefore, $D_1\cap D_{p+1}=\varnothing$.
To prove ( d ), use the induction. For n=1,2, the proposition is checked directly (fig. 6.3, 6.4). Assume that the proposition is valid whenever n<p. Let n=p. Suppose that $V=\{D_2,D_3\ldots D_m\}$ is the collection of n-disks intersecting $D_1$. Then $U=\{E_2,E_3\ldots E_m\}$, $E_k=D_k\cap D_1$ is the LCL cover of an (n-1)-sphere $\partial D_1$ by (n-1)-disks according to proposition 6.3. Then the amount x of elements in U is more than or equal to 2n, $x\geq 2n$ by the assumption. Since there is at least one n-disk not intersecting $D_1$ (proposition 6.4(a)), then $t\geq 2n+2$. This completes the proof. □

Proposition 6.5.
Suppose that collection $W=\{D_0,D_1,D_2\ldots D_t\}$ of n-disks is an LCL cover of an n-sphere S and $V=\{D_1,D_2\ldots D_p\}$ is a collection of all n-disks intersecting $D_0$. Then collection $U=\{D_0,D_1,D_2\ldots D_p,C\}$, where $C=D_{p+1}\cup D_{p+2}\cup\ldots D_t$, is an LCL cover of S by n-disks such that if $D_{i(1)}\cap D_{i(2)}\cap D_{i(m)}\neq\varnothing$, $D_{i(k)}\in V$, $k=1,2,\ldots m$, then $C\cap D_{i(1)}\cap D_{i(2)}\cap D_{i(m)}\neq\varnothing$.
Proof.
Obviously, U is a cover of S. According to proposition 6.3, the union $A=D_0\cup D_1\cup D_2\cup\ldots D_p$ is an n-disk. Hence, $C=S-\text{Int}A$ is an n-disk. By proposition 6.4, $H_i=C\cap D_i\neq\varnothing$, $i=1,2,\ldots p$.
For n=1,2, the proposition is verified directly. Assume that the proposition is valid whenever $n\leq s$. Let $n=s+1$. Suppose that $X_1=\{D_0,D_2,D_3,D_4\ldots D_m,D_{p+1},D_{p+2}\ldots D_{p+h}\}$ is the collection of all n-disks belonging to W and intersecting $D_1$, where $D_i\in V$, $i=2,3,\ldots m$, $D_i\notin V$, $i=p+1,p+2,\ldots p+h$. As in the proof of proposition 6.4(b), h>0 and $Y_1=\{E_0,E_2,E_3,\ldots E_m,E_{p+1},E_{p+2}\ldots E_{p+h}\}$, $E_i=D_2\cap D_i$, $D_i\in X_1$, is an LCL cover of an (n-1)-sphere $\partial D_2$ by (n-1)-disks according to proposition 6.3. Then collection $Z_1=\{E_0,E_2,E_3\ldots E_m,H_1\}$, where $H_1=E_{p+1}\cup E_{p+2}\cup\ldots E_{p+h}=C\cap D_1$, is an LCL cover of $\partial D_1$ by (n-1)-disks according to the assumption. Since $D_1$ is taken arbitrarily, then $Z_k$, k=1,2,…p, is an LCL cover of $\partial D_k$ by (n-1)-disks. Obviously, $Z_C=\{H_1,H_2,\ldots H_m\}$ is a cover of $\partial C$.

First we have to show that $U=\{D_0,D_1,D_2\ldots D_p,C\}$ is locally centered. Suppose that $D_i\cap D_k\neq\varnothing$, i,k=1,2,…f. Then $E_2\cap E_3\cap\ldots E_f=D_1\cap D_2\cap D_3\cap\ldots D_f\neq\varnothing$. Since $E_i\in Z_1$, i=2,3,…f, and $D_i\cap Z_1\neq\varnothing$, then $E_2\cap E_3\cap\ldots E_f\cap H_1\neq\varnothing$ according to the assumption. Hence, $C\cap D_1\cap D_2\cap D_3\cap\ldots D_f=E_2\cap E_3\cap\ldots E_f\cap H_1\neq\varnothing$ and U is locally centered. The intersection $E_2\cap E_3\cap\ldots E_f\cap H_1=B$ is an (n-f)-disk by the assumption. Since $C\cap D_1\cap D_2\cap D_3\cap\ldots D_f=B$, then U is LCL collection according to definitions 6.1 and 6.3. Hence, U is an LCL cover of S such that if $D_{i(1)}\cap D_{i(2)}\cap D_{i(m)}\neq\varnothing$, $D_{i(k)}\in V$, k=1,2,…m, then $C\cap D_{i(1)}\cap D_{i(2)}\cap D_{i(m)}\neq\varnothing$. The proof is complete. □

Further for technical convenience, let us call the collection of sets $W=\{u_1,u_2,\ldots\}$ contractible, if the intersection graph G(W) of W is contractible, let us call the rim of $u_k$ the collection $O(u_k)$ of all sets belonging to W and intersecting $u_k$.

Proposition 7.1.
Suppose that $W=\{u_1,u_2,\ldots\}$ is a tiling of the n-dimensional Euclidean space $R^n$ into a family of n-cubes with the edge length L, B is an n-box in $R^n$, $U=\{u_1,u_2,\ldots u_s\}$ is a family of n-cubes intersecting B. Then the intersection graph G(U) of U is contractible.
Proof.
Obviously, U is a cover of B. For small number s, it is checked directly. With no loss of generality, suppose that the edges of B are parallel to the coordinate axes, L is much smaller than the length of the shortest edge r of B, L<<r, and if $B\cap u_k\neq\varnothing$, then $IntB\cap Intu_k\neq\varnothing$ (fig. 10.1).
Let $U=\{u_{11\ldots1},\ldots u_{mp\ldots q}\}$. Obviously, for any cube $u_{1a\ldots b}$ there is a cube $u_{2a\ldots b}$ such that $u_{2a\ldots b}$ is adjacent to all other cubes belonging to the rim $O(u_{ma\ldots b})$. Therefore, the rim $O(u_{1a\ldots b})$ is contractible and all cubes $u_{1a\ldots b}$ can be deleted. In the same way, all cubes $u_{2a\ldots b}, u_{3a\ldots b}, \ldots u_{ma\ldots b}$ can be deleted except for cube $u_{mp\ldots q}$. The proof is completed. □

Note that G(U)=G(V), where $V=\{e_1,e_2,\ldots e_s\}$ is a cover of B by $e_k=B\cap u_k$.

Proposition 7.2.
Suppose that $W=\{u_1,u_2,\ldots\}$ is a tiling of the n-dimensional Euclidean space $R^n$ into a family of n-cubes with the edge length L, D is a finite convex n-disk in $R^n$, $U=\{u_1,u_2,\ldots u_s\}$ is a family of n-cubes intersecting D. Then the intersection graph G(U) of U is contractible.

Proof.
To simplify the proof, consider the dimension two (fig 10.2). Suppose that a point (x,y) belongs to the cube $u_{kp}$ if $x_0+kL\leq x\leq x_0+(k+1)L$, $y_0+pL\leq y\leq y_0+(p+1)L$, $k\in Z$. Let $U=\{u_{kp}\}$ be the cover of a convex finite two-disk D such that for any $u_{kp}\in U$, the intersection $e_{kp}=D\cap u_{kp}$ is a closed n-disk.
Denote $X_k$ the collection of cubes belonging to cover U whose first coordinate equal to k and denote $Y_p$ the collection of cubes belonging to cover U whose second coordinate equals p.
Call $X_k$ the boundary level if $X_{k+1}$ (or $X_{k-1}$) is empty.

Suppose that $Y_{p+1}$ is empty, $Y_p$ is not empty and for any $u_{kp} \in Y_p$ there is $u_{k,p-1} \in Y_{p-1}$. Then the set of cubes adjacent to $u_{kp}$ is contractible and any $u_{kp}$ can be deleted. Obviously, $U_p = U - Y_p$ is the cover of the convex two-disk $D_p = D - Int|Y_p|$.
Suppose that $Y_{p+1}$ is empty, $Y_p$ is not empty and there is $u_{kp} \in Y_p$ such that $u_{k,p-1}$ does not belong to the cover, $u_{k,p-1} \notin Y_{p-1}$. Assume that there is some $u_{k,p-s}$ belonging to the cover, $u_{k,p-s} \in U$. Then there are points a and b such that $a \in Int(u_{kp})$, $b \in Int(u_{kp-1})$. Since D is convex, then the line segment [a,b] must intersect cube $u_{k,p-1}$. It contradicts the assumption. Therefore, $u_{k,p-s} \notin U$ for any s. If $u_{k+1,p} \notin U$, then $X_k$ is the boundary level containing the only $u_{kp}$ which can be deleted. If $u_{k+1,p} \in U$, then there is $u_{k+m,p} \in U$, $m \geq 0$, such that $u_{k+m+1,p} \notin U$. Obviously, $u_{k+m,p}$ belongs to $X_{k+m}$ and can be deleted. Therefore, for any cover, we have the boundary level, say $A_p$, which can be deleted. Delete from U any $u_{kp} \in Y_{p-1}$ and delete from D points belonging to $D \cap Int(u_{kp})$. We obtain the cover $U_1$ of the convex closed two-disk $D_1$. Obviously, we can apply this procedure to $U_1$ and $D_1$. Finally, we convert it to one cube and G(U) to one point. In the same way, the proposition can be proven for dimension n>2. □

Proposition 7.3.
Let $W = \{H_1, H_2, \ldots H_t\}$ be a locally centered collection of convex finite closed n-polytopes such that if $H_i \cap H_k \neq \varnothing$, then $H_{ik} = H_i \cap H_k$ is an (n-$p_{ik}$)-polytope, $0 < p_{ik} \leq n$. Suppose that $F = \{u_1, u_2, \ldots\}$ is a tiling of the m-dimensional Euclidean space $R^m$ (m≥n) into a family of m-cubes with the edge length L and $U = \{u_1, u_2, \ldots u_q\}$ is a family of m-cubes intersecting $A = H_1 \cup H_2 \cup \ldots H_t$. If the intersection graph G(W) of W is contractible, then there is r>0, such that for any L<r, the intersection graph G(U) of U is contractible.
Proof.
The proof is by induction. For dimension n=1, the proposition is verified directly (fig 10.3). Assume that the proposition is valid whenever n<a+1. Let n=a+1.
Suppose that t=2. According to proposition 7.2, there is r>0, such that for any L<r, the intersection graphs $G(U_1)$ and $G(U_2)$ of collections $U_1$ and $U_2$ of cubes intersecting $H_1$ and $H_2$ are contractible. $H_{12} = H_1 \cap H_2$ is a convex closed (n-p)-polytope. Therefore, by proposition 7.2, there is $r_{12} > 0$, such that for any $L < r_{12}$, the intersection graph $G(U_{12})$ of collection $U_{12}$ of cubes intersecting any $H_{12}$ is contractible. Suppose that d is the minimum of r and $r_{12}$ and L<d. Then $G(U_1)$, $G(U_2)$ and $G(U_{12})$ are contractible $G(U) = G(U_1) \cup G(U_2)$ and $G(U_{12}) = G(U_1) \cap G(U_2)$. By proposition 3.4, $G(U_1)$ can be converted into $G(U_{12})$ by contractible transformations. Therefore, G(U) can be converted into $G(U_2)$ by the same transformations. Since $G(U_2)$ is contractible, then it can be converted to a point.
Assume that the proposition is valid whenever t<b+1. Let t=b+1.
Since G(W) is contractible, then there is $H_k \in W$, say $H_1$, such that the intersection graph $O(W_1)$ of the family $W_1$ of n-polytopes intersecting $H_1$ is contractible. Note that $H_1$ does not belong to $W_1$. Suppose $W_1 = \{H_2, H_3, \ldots H_q\}$. Then $V_1 = \{H_{21}, H_{31}, \ldots H_{q1}\}$ is a locally centered collection of convex finite closed (n-$p_{i1}$)-polytopes, $p_{i1} > 0$. Obviously, the intersection graph $G(V_1)$ of $V_1$ is isomorphic to $G(W_1)$ and, therefore, $G(V_1)$ is contractible.
Consider $H_1$, $B_1 = H_{21} \cup H_{31} \cup \ldots H_{q1}$ and $E = H_2 \cup H_3 \cup \ldots H_t$. By proposition 7.3 and the first and the second assumptions, there is r>0, such that such that for any L<r, the intersection graphs $G(U_1)$, $G(U_B)$ and $G(U_E)$ of collections $U_1$, $U_B$ and $U_E$ of cubes intersecting any $H_1$ and $B_1$ and E are contractible. Obviously, $G(U) = G(U_1) \cup G(E)$ and $G(B_1) = G(U_1) \cap G(E)$. By proposition 3.4, $G(U_1)$ can be converted into $G(B_1)$ by contractible transformations. Therefore, G(U) can be converted into G(E) by the same transformations. Since G(E) is contractible by the second assumption, then it can be converted to a point. The proof is completed. □

Proposition 7.4.
Suppose that $W=\{u_1,u_2,...\}$ is the tiling of the n-dimensional Euclidean space $R^n$ into a family of n-cubes with the edge length L, D is a finite closed n-disk in $R^n$, $U=\{u_1,u_2,...u_t\}$ is a family of n-cubes intersecting D. Then there is r>0 such that for any L<r, the intersection graph G(U) of U is contractible (fig. 10.4).
Proof.
Suppose that f is a homeomorphism from $R^n$ onto $R^n$ such that Y=f(D) is an n-box, $P=\{p_1,p_2,...\}$ is a tiling of the n-dimensional Euclidean space $R^n$ into a family of n-cubes with the edge length M, $Q=\{p_1,p_2,...p_s\}$ is a family of n-cubes intersecting Y, and $Z=\{a_1,a_2,...a_s\}$ is the collection of n-boxes such that $a_k=p_k \cap Y$. Obviously, collections Q and Z are locally centered and collection Z is a cover of Y such that $a_1 \cup a_2 \cup ... \cup a_s = Y$. Note that the intersection graph G(Z) of Z is isomorphic to the intersection graph G(Q) of Q. By proposition 7.1, G(Z) is contractible. By construction, $Z=\{a_1,a_2,...a_s\}$ is a locally centered collection of convex finite closed n-polytopes such that if $a_i \cap a_k \neq \varnothing$, then $a_{ik}=a_i \cap a_k$ is an (n-$p_{ik}$) polytope, $p_{ik}>0$.
It is clear that collection $B=\{f^{-1}(a_1),f^{-1}(a_2),...f^{-1}(a_s)\}$ of inverse images is a locally centered cover of D by n-disks $f^{-1}(a_i)$ with properties similar to properties of cover Z. If M is sufficiently small, then shapes of all $f^{-1}(a_i)$ are close to the shapes of convex n-polytopes $e_i$ with vertices $f^{-1}(x_{ik})$ where $x_{ik}$ are vertices of $a_i$. Therefore, we can replace all $f^{-1}(a_i)$ by $e_i$ and $W=\{e_1,e_2,...e_s\}$ is a locally centered collection of convex finite closed n-polytopes such that if $a_i \cap a_k \neq \varnothing$ and $a_{ik}=a_i \cap a_k$ is an (n-$p_{ik}$)-polytope, $p_{ik}>0$, then $e_i \cap e_k \neq \varnothing$ and $e_{ik}=a_i \cap e_k$ is an (n-$p_{ik}$)-polytope.
The intersection graph G(W) of W is isomorphic to G(Z) and, therefore, contractible. By proposition 7.3, there is a tiling of $R^n$ into a family $F=\{u_1,u_2,...\}$ of n-cubes with the edge length L and r>0 such that for any L<r, the intersection graph G(U) of family $U=\{u_1,u_2,...u_q\}$ of n-cubes intersecting $E=e_1 \cup e_2 \cup ... e_s$ is contractible. Note that approximation of D by E can get arbitrarily close to D. The proof is completed. □

Notice that the tiling of the n-dimensional Euclidean space $R^n$ into the family of n-cubes with the edge length L is not the only one, which can be used to prove previous results. In fact, we can use a wide range of tesselations and covers of $R^n$. Denote by d(e) the maximal distance between pairs of points in an n-disk e. It is not difficult to prove the following corollary.

Corollary 7.1.
Suppose that $W=\{e_1,e_2,...\}$ is a cover of the n-dimensional Euclidean space $R^n$ by a family of n-disks such that the diameter $d(e_k)$ of any n-disk $e_k$ is smaller than L. Suppose that for any n-cube C there is r>0 such that for L<r, the intersection graph G(V) of the family $V=\{e_1,e_2,...e_s\}$ of n-disks intersecting C is contractible. Then for any n-disk D there is d>0 such that for L<d, the intersection graph G(U) of the family $U=\{e_1,e_2,...e_t\}$ of n-disks intersecting D is contractible.

Proposition 7.5.
Let $W=\{D_1,D_2,...D_s\}$ be an LCL collection of n-disks and $P=D_1 \cup D_2 \cup ... D_s$ be an n-disk. Then the intersection graph G(W) of W is contractible.

Proof.
Since W contains a finite amount of closed n-disks, then we can always find a collection $U=\{B_1, B_2,\ldots B_s\}$ of n-disks with the following properties (fig. 10.5(a,b)).
$D_k \subseteq IntB_k$, $k=1,2,\ldots s$.
$B_i \cap B_k \cap \ldots B_p \neq \varnothing$ if and only if $D_i \cap D_k \cap \ldots D_p \neq \varnothing$.
Obviously, U is a locally centered collection such that if $B_i \cap B_k \cap \ldots B_p$, then $F=B_i \cap B_k \cap \ldots B_p$ is a closed n-disk.
Let $C_k=B_k \cap D$, $k=1,2,\ldots s$. Obviously, collection $V=\{C_1, C_2,\ldots C_s\}$ is a locally centered collection of n-disks and V is the cover of D with the following properties (fig. 10.5(c)).
$D_k \subseteq C_k$, $k=1,2,\ldots s$, and $D=C_1 \cup C_2 \cup \ldots \cup C_s$.
$C_i \cap C_k \cap \ldots C_p \neq \varnothing$ if and only if $D_i \cap D_k \cap \ldots D_p \neq \varnothing$.
if $C_{k(m)} \cap C_{k(n)} \neq \varnothing$, $m,n=1,2,\ldots p$, then $F=B_i \cap B_k \cap \ldots B_p$ is a closed n-disk.
If $D_i \cap D_k \cap \ldots D_p \neq \varnothing$ then $F_{ik\ldots p}=C_i \cap C_k \cap \ldots C_p$ is a closed n-disk.
By construction, the intersection graph G(W) of W is isomorphic to the intersection graph G(V) of V.
Suppose that $X=\{u_1, u_2,\ldots\}$ is a tiling of the n-dimensional Euclidean space $R^n$ into a family of n-cubes with the edge length L, U is a family of n-cubes intersecting D, $U_k$ is a family of cubes intersecting $C_k$, $k=1,2,\ldots s$ and $U_{ik\ldots p}$ is a family of cubes intersecting a non-empty n-disk $F_{ik\ldots p}$,
By proposition 7.4, for D, there is $d>0$ such that for any $L<d$, the intersection graphs G(U), $G(C_k)$ and $G(F_{ik\ldots p})$ of collections of cubes intersecting D, all $C_k$ and all $F_{ik\ldots p}$ are contractible.
Since any $D_k$ belongs to $D \cap IntC_k$, then for any $D_k$, there is $d_k>0$, such that if $L<d_k$, then the following condition holds (fig. 10.5(d)). If $D_k \cap u_p \neq \varnothing$, then for any $u_i$ such that $u_i \cap u_p \neq \varnothing$ and $D \cap u_i \neq \varnothing$, it follows that $C_k \cap u_i \neq \varnothing$.
Suppose that r is the minimum of d and $d_k$, $k=1,2,\ldots s$, and $L<r$.
Then:
Any $G(U_k)$, $k=1,2,\ldots s$, is contractible.
if $G(U_i) \cap G(U_k) \cap \ldots \cap G(U_p) \neq \varnothing$, then $G(U_i) \cap G(U_k) \cap \ldots \cap G(U_p) = G(U_{ik\ldots p})$ is contractible.
If $D_k \cap u_p \neq \varnothing$, then the intersection graph $G(Z_p)$ of collection $Z_p$ of cubes adjacent to $u_p$ and intersecting D belongs to $G(U_p)$. Then the intersection graph G(Y) of collection $Y=\{G(U_1), G(U_2),\ldots G(U_s)\}$ is homotopic G(U) according to proposition 3.5. Since G(U) is contractible, then G(Y) is contractible. Obviously, G(Y) is isomorphic to G(V) and G(V) is isomorphic to G(W). Hence, G(W) is contractible. □

Proposition 7.6.
Let $W=\{D_1,\ldots D_s\}$ be an LCL collection of n-disks and $P=D_1 \cup D_2 \cup \ldots D_s$. If the intersection graph G(W) of W is contractible, then P is an n-disk.
Proof.
For small numbers s it is checked directly (fig. 7.1). Assume that the proposition is valid whenever $s<p+1$. Let $s=p+1$. Suppose that G(W) contains points $\{v_1, v_2,\ldots v_s\}$ and the rim $O(v_1)$ of point $v_1$ is contractible. Delete point $v_s$ from G(W). Since subgraph $G_1=G(W)-v_1$ is also contractible and it is the intersection graph of collection $W=\{D_2,\ldots D_s\}$, then $B=D_2 \cup D_3 \cup \ldots D_s$ is an n-disk by the assumption. Suppose that the rim of point $v_1$ contains points $v_2, v_3,\ldots v_m$. Then the collection $\{C_2, C_3,\ldots C_m\}$ of (n-1)-disks $C_k=D_1 \cap D_k$, $k=2,3,\ldots m$, is the LCL collection. Therefore, the union $C=C_2 \cup C_3 \cup \ldots C_m$ is an (n-1)-disk by the assumption. Therefore $P=D_1 \cup B$ is the union of two n-disks such that their intersection $C=D_1 \cap B$ is an (n-1)-disk. Therefore, P is an n-disk. □

The following statement is an easy consequence of propositions 7.5 and 7.6.

Theorem 7.1.

Let $W=\{D_1,\ldots D_s\}$ be an LCL collection of n-disks and $P=D_1 \cup D_2 \cup \ldots D_s$. The intersection graph G(W) of W is contractible if and only if P is an n-disk (fig. 7.1).

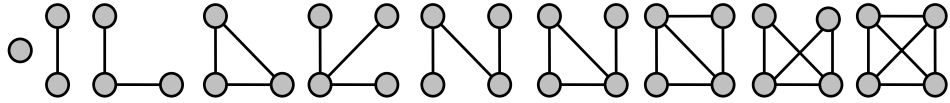

Figure 3.1. Contractible graphs with a number of points n<5.

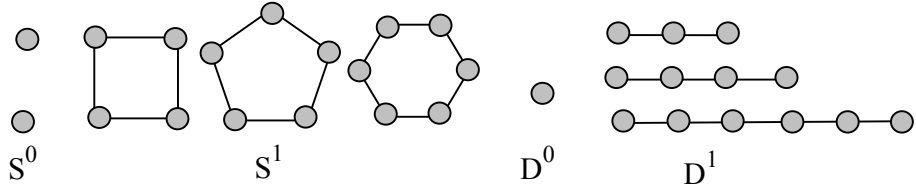

Figure 4.1. Zero- and one-spheres $S^0$ and $S^1$ and zero- and one-disks $D^0$ and $D^1$.

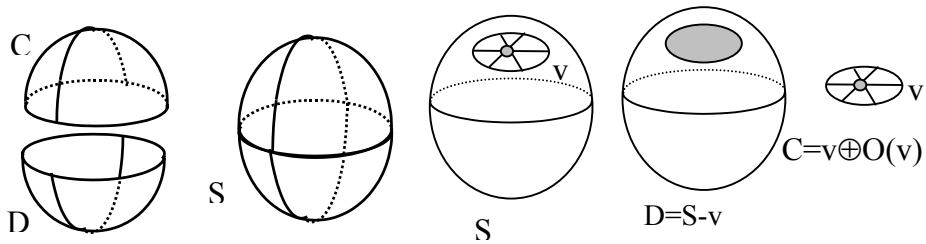

Figure 4.2. An n-sphere S is the connected sum of n-disks C and D. If S is an n-sphere, then D=S-v is an n-disk

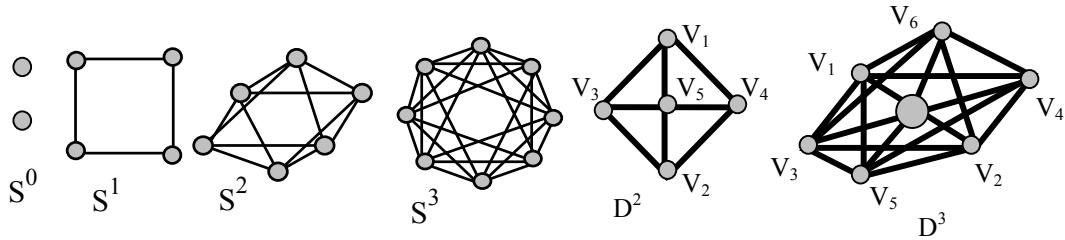

Figure 4.3. Minimal spheres and disks.

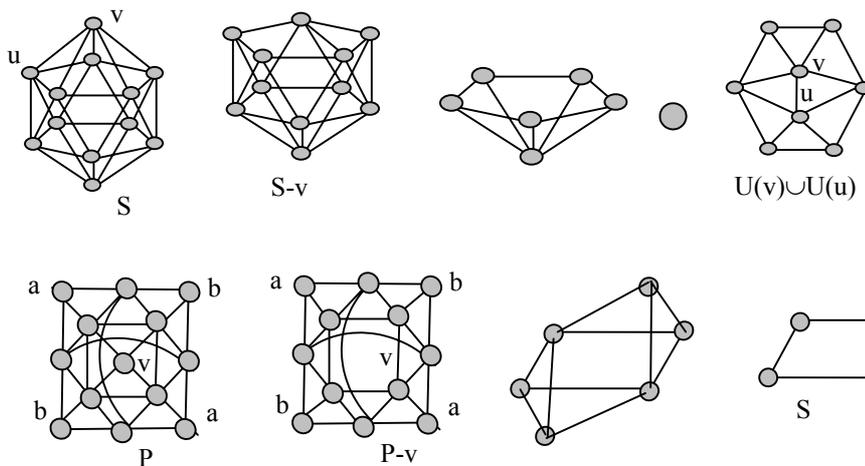

Figure 4.4. S is a 2-sphere, S-v is a 2-disk, which is homotopic to a point. S is not compressed. The union U(v)∪U(u) of balls is a two-disk. P is a 2-dimensional projective plane, P-v is homotopic to a 1-sphere S.

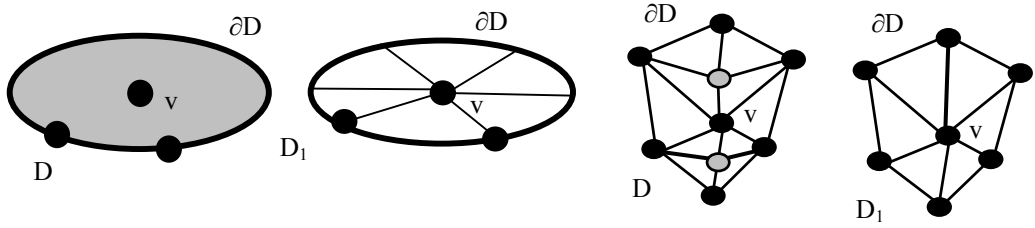

Figure 4.5. The replacing of n-disk D by n-disk $D_1=v \oplus \partial D$. In $D_1$, point v is adjacent to any point of the boundary $\partial D$.

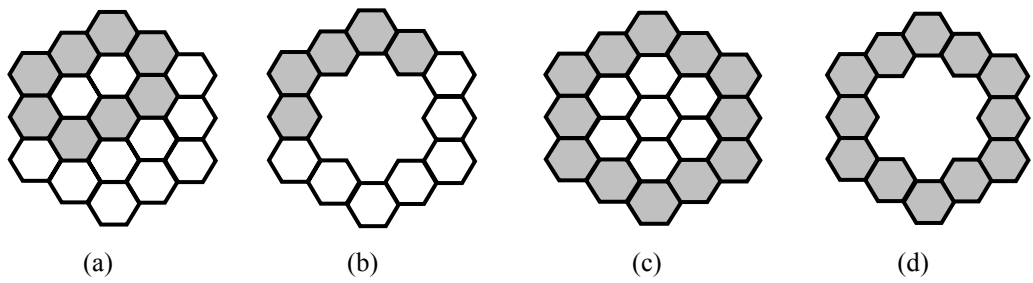

Figure 4.6. (a), (c). Segmented 1-spheres S (gray). (b). S is collapsed by the merging of all interior elements of a segmented 2-disk. (d). After the merging of all interior elements of a segmented 2-disk, S becomes the boundary of a segmented 2-disk with one interior element.

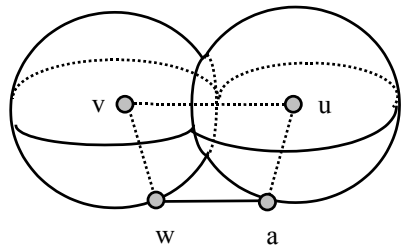

Figure 5.1. In a compressed n-manifold, any two adjacent points v and u belong to a 1-sphere {v,u,a,w} containing four points.

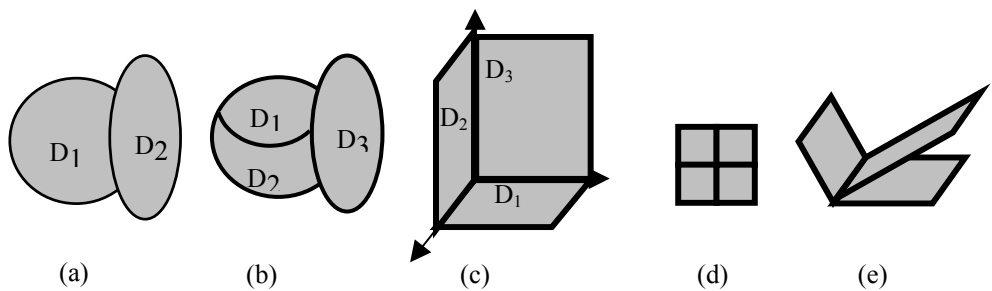

Figure. 6.1. (a), (b) and (c) are lump collections. (d) and (e) are not lump collections.

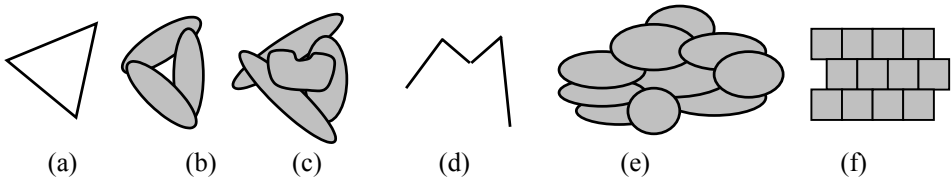

Figure 6.2. Collections (a), (b) and (c) are not locally centered but contain lump subcollections. Collections (d), (e) and (f) are locally centered lump collections of 1- and 2-dimensional disks.

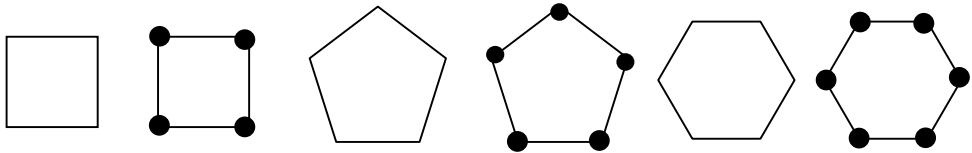

Figure. 6.3. LCL covers of a circle and their intersection graphs. Digital models of a circle are digital 1-spheres.

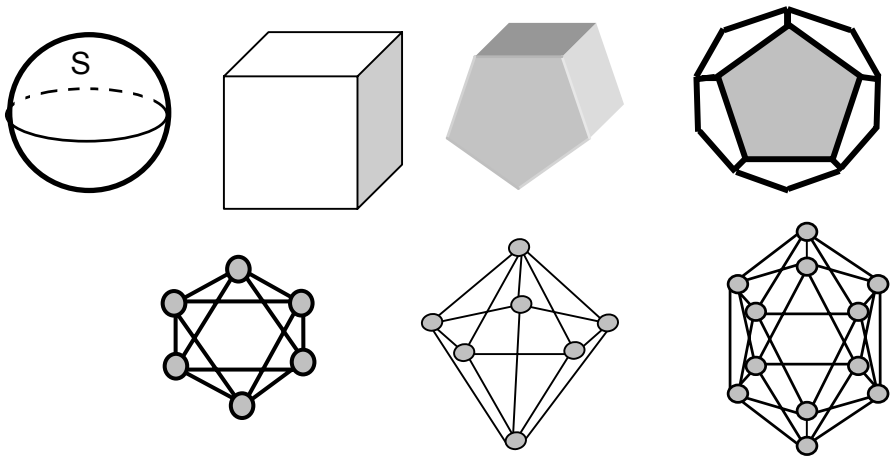

Figure. 6.4. LCL covers of a continuous 2-sphere S and their digital models. The minimal LCL cover contains six 2-disks.

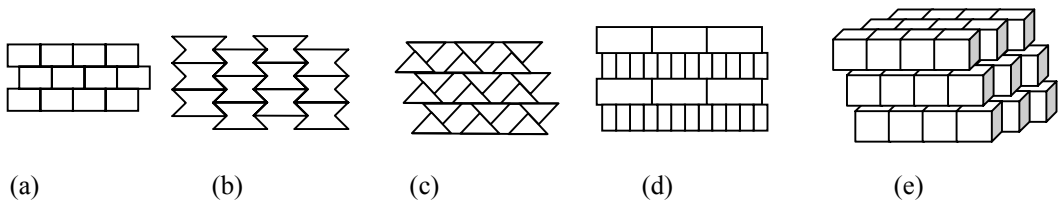

Figure. 6.5. (a), (b), (c), (d) are LCL tiling of a 2-plane. Tiling (e) is an LCL tessellation of Euclidean 3-space.

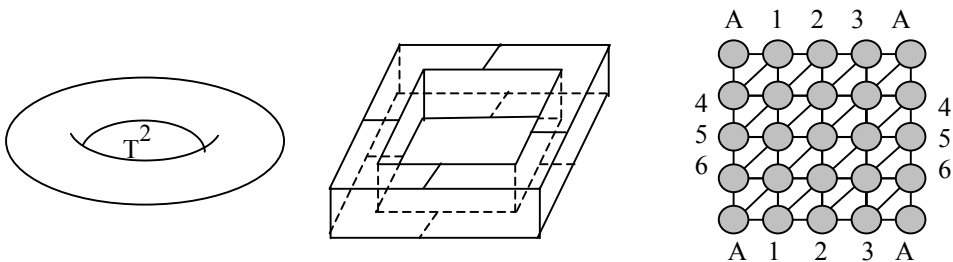

Figure. 6.6. An LCL cover of the continuous 2-dimensional torus $T^2$. The intersection graph of this cover is the minimal digital 2-torus.

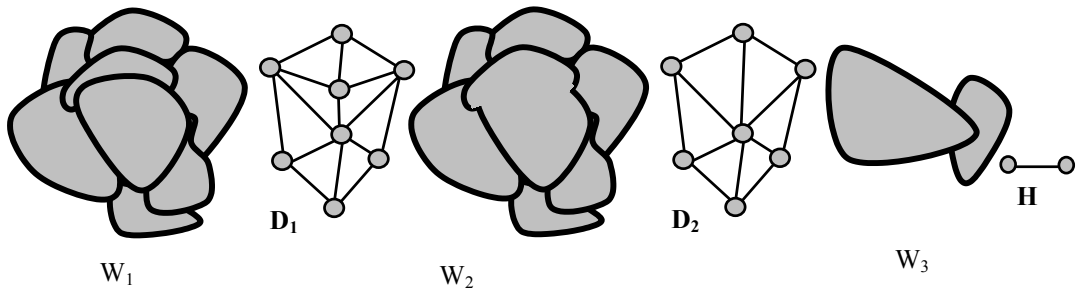

Figure 6.7. $W_1$ and $W_2$ are segmented 2-disks. Collection $W_3$ is not a segmented 2-disk.

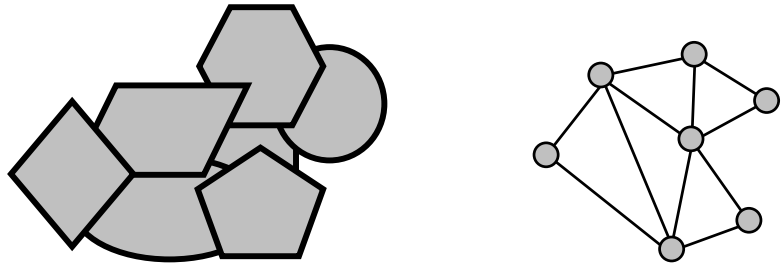

Figure 7.1. An LCL collection of n-disks is contractible if and only if the union of these disks is an n-disk.

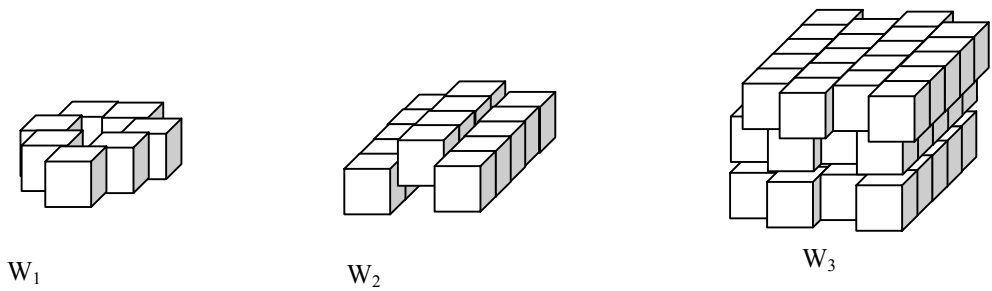

Figure 7.2. $W_1$ is a segmented 1-sphere, $W_2$ is a segmented 2-disk, $W_3$ is a segmented 3-disk.

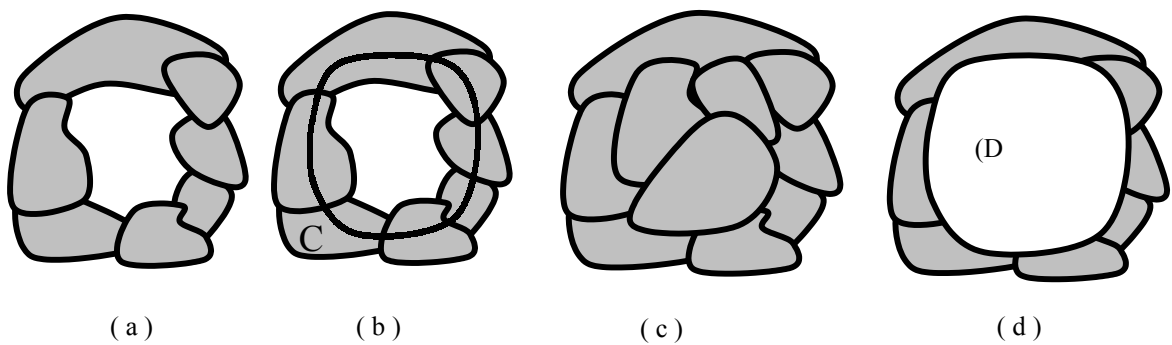

Figure 7.3. (a). A segmented 1-sphere S. (b). A closed curve C is a continuous analog of S. (c). A segmented 2-disk U. (d). A continuous disk D is continuous analog of U.

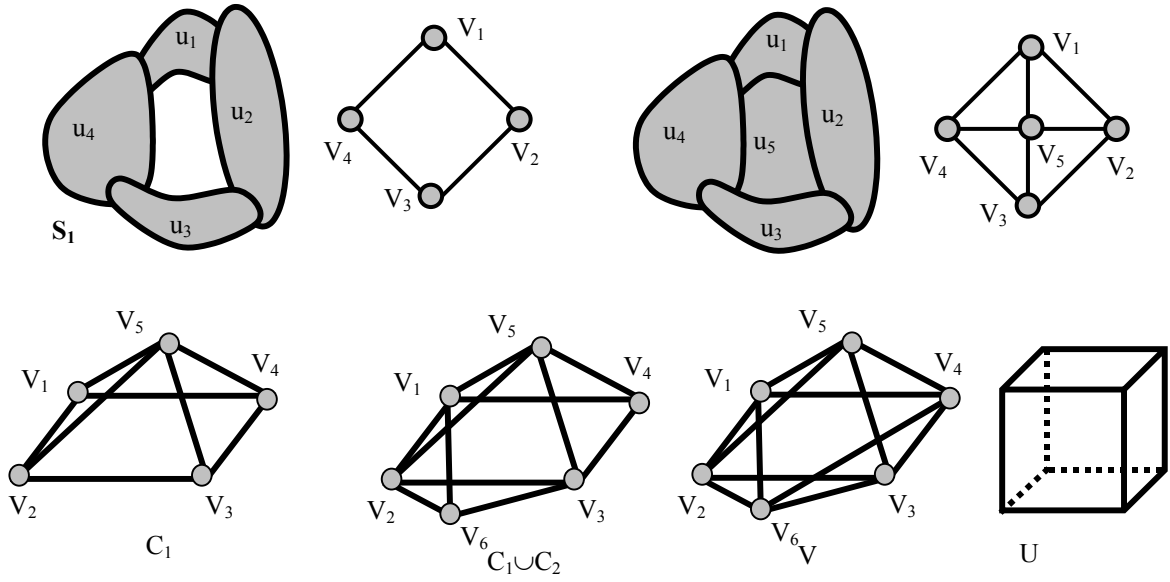

Figure 8.1. An irreducible LCL cover of M isomorphic to the collection of two-dimensional faces of the three-dimensional unit cube U.

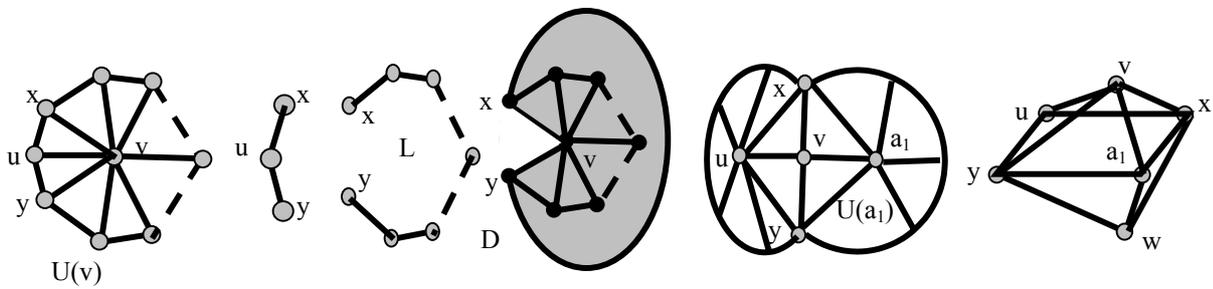

Figure 8.2. The rim of point v is the minimal 1-sphere $S(4)=\{x,a_1,y,u\}$ and G is the minimal 2-sphere $S^2(6)=\{v,x,a_1,y,u,w\}$.

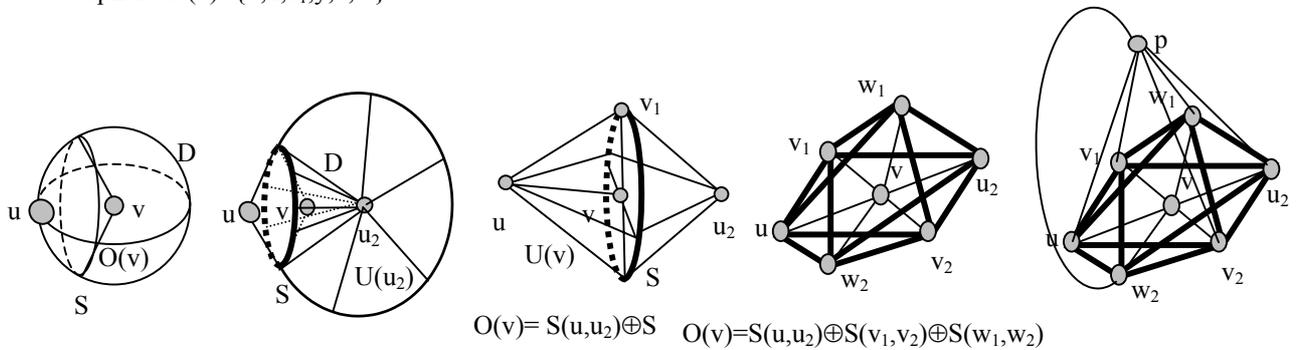

Figure 9.1. A digital 2-disk D belongs to a digital 3-disk $U(u_2)$. $U(v)$ is the ball of point v. $S^3_{min}=S(u,u_2)\oplus S(v_1,v_2)\oplus S(w_1,w_2)\oplus S(v,p)$ is the minimal digital 3-sphere.

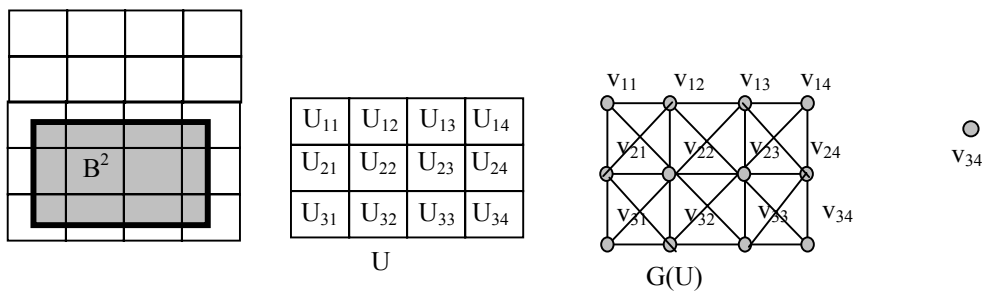

Figure 10.1. The two-box $B^2$, its cover U and the intersection graph G(U) of U. G(U) can be converted to a point $v_{34}$ by the deleting of points $v_{11}, v_{12},...v_{33}$.

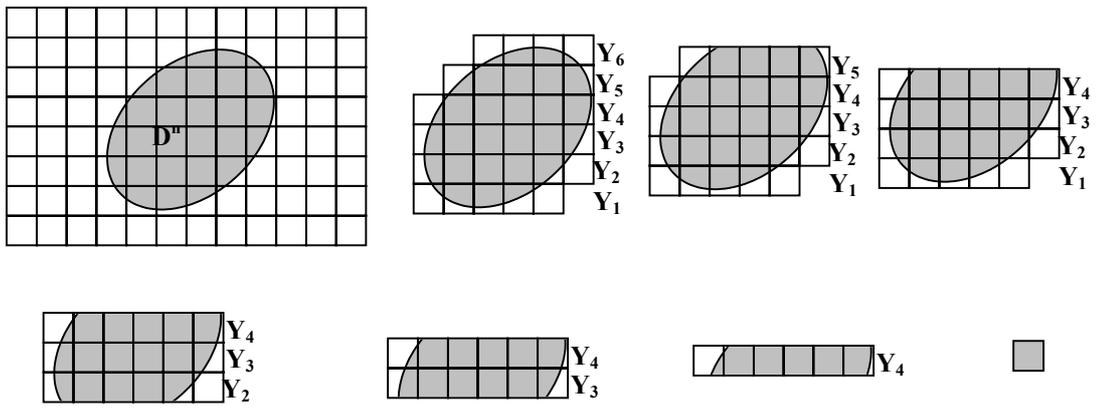

Figure 10.2. The digital model G(U) of a convex n-disk $D^2$ can be converted into a point by the deleting of cubes belonging to layers $Y_6$, $Y_5$, $Y_1$, $Y_2$ and $Y_3$.

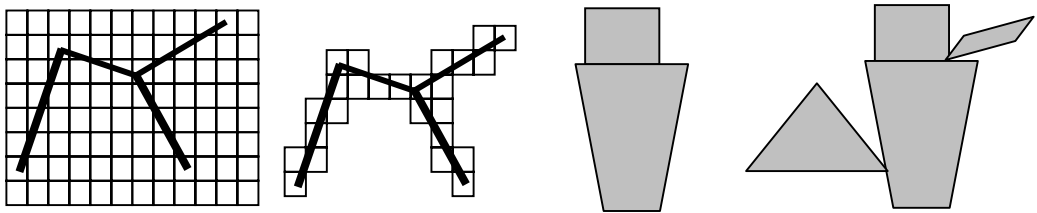

Figure 10.3. Contractible collections of one and two-dimensional convex polytopes.

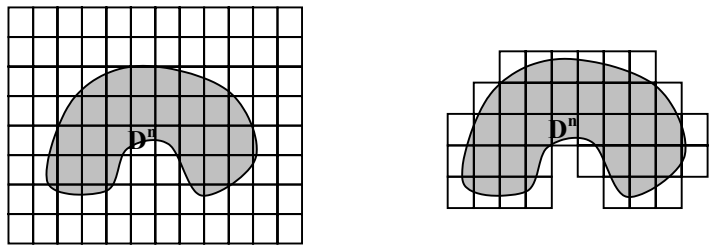

Figure 10.4. The digital model of an n-disk $D^n$ is a contractible graph.

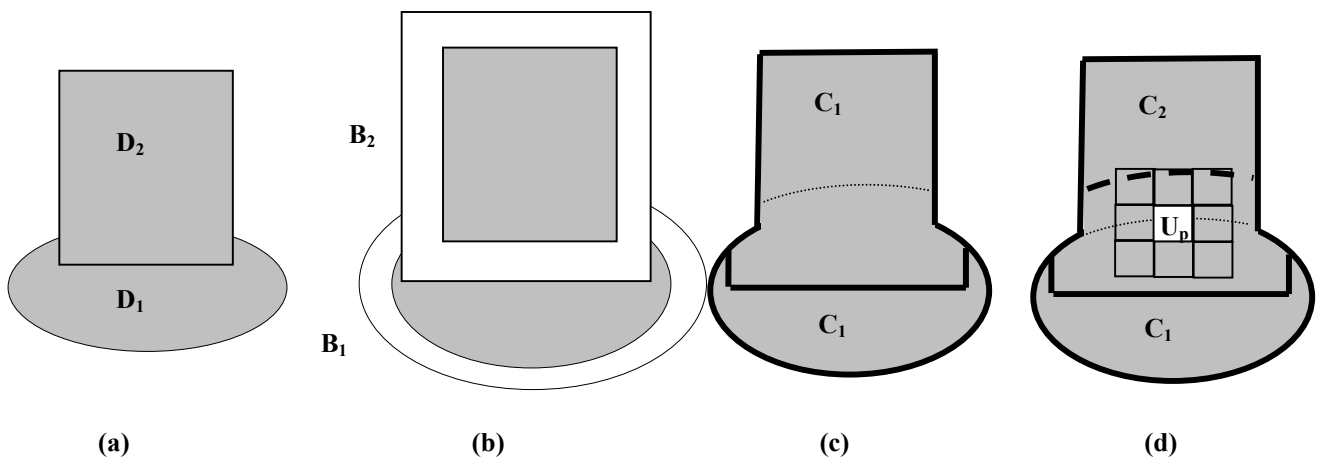

Figure 10.5. An LCL collection {$D_1$,$D_2$} is replaced by a locally centered collection {$C_1$,$C_2$}.